\documentclass[letterpaper, 10 pt, conference]{ieeeconf}  

\IEEEoverridecommandlockouts                              

\usepackage{graphics} 
\usepackage{subfigure}
\usepackage{epsfig} 
\usepackage{multicol}
\usepackage{amssymb}  
\usepackage{algorithm}
\usepackage{algorithmicx}
\usepackage{algpseudocode}
\usepackage{soul}
\usepackage{comment}
\usepackage{epstopdf}
\usepackage{hyperref}
\usepackage{multirow}
\usepackage{cite}
\usepackage{soul,xcolor}
\usepackage{todonotes}
\usepackage{amsmath}
\usepackage{hyperref}
\usepackage{array}
\usepackage{balance}
\usepackage[flushleft]{threeparttable}

\newtheorem{lemma}{Lemma}

\newtheorem{definition}{Definition}

\newcommand{\tr}{\mathsf{ T}}

\title{\LARGE \bf
	Data-Enabled Predictive Control for Flexible Spacecraft}   

\author{
    Huanqing Wang,
    Kaixiang Zhang,
    Amin Vahidi-Moghaddam,
    Haowei An,
    Nan Li,
    Daning Huang,
    Zhaojian Li
    \thanks{This work was partially supported by the U.S. National Science Foundation projects  CMMI-2320698. \textit{(Corresponding author: Zhaojian Li.)}}
    \thanks{Huanqing Wang, Kaixiang Zhang, Amin Vahidi-Moghaddam, Haowei An and Zhaojian Li are with the Department of Mechanical Engineering, Michigan State University, East Lansing, MI 48824, USA (e-mail: \{wanghu26, zhangk64, vahidimo, anhaowei, lizhaoj1\}@egr.msu.edu).}%
    \thanks{Nan Li is with the School of Automotive  Studies, Tongji University, Shanghai, China (e-mail: li\_nan@tongji.edu.cn).}%
    \thanks{Daning Huang is with the Department of Aerospace Engineering, Pennsylvania State University, University Park, PA 16802, USA (e-mail: daning@psu.edu).}
}

\begin{document}
	
	\maketitle
	\thispagestyle{empty}
	\pagestyle{empty}

\begin{abstract}
		
Spacecraft are vital to space exploration and are often equipped with lightweight, flexible appendages to meet strict weight constraints. These appendages pose significant challenges for modeling and control due to their inherent nonlinearity. Data-driven control methods have gained traction to address such challenges. This paper introduces, to the best of the authors' knowledge, the first application of the data-enabled predictive control (DeePC) framework to  boundary control for flexible spacecraft. 
Leveraging the fundamental lemma, DeePC constructs a non-parametric model by utilizing recorded past trajectories, eliminating the need for explicit model development. 
The developed method also incorporates dimension reduction techniques to enhance computational efficiency. 
Through comprehensive numerical simulations, this study compares the proposed method with Lyapunov-based control, demonstrating superior performance and offering a thorough evaluation of data-driven control for flexible spacecraft.

\end{abstract}
	
\textit{\textbf{Index Terms}}---Flexible spacecraft, data-driven control, predictive control.

	\section{Introduction}
	
Flexible spacecraft have attracted considerable interest due to their wide range of applications, including space missions, communication, and remote sensing \cite{Yin2016}. The high cost of launches has driven the demand for lighter structures and compact servomechanisms. To minimize weight, spacecraft components such as solar panels and antennas are often designed as lightweight and flexible structures. Spacecraft such as Voyager 1 feature long, extended appendages that position instruments, such as magnetometers, farther from the spacecraft to reduce electromagnetic interference. These long structures, often equipped with sensors, are oriented at specific angles to ensure accurate measurements or signal transmission. However, this flexibility induces vibrations that can impair functionality and performance. Prolonged undesirable vibrations can reduce the spacecraft's lifespan and even lead to failure.

For complex systems like flexible spacecraft, the dynamics are often represented by partial differential equations (PDEs) coupled with ordinary differential equations (ODEs) \cite{Hu2007,HU2005}. Various control methods have been  investigated for such systems, including proportional-derivative (PD) control \cite{Baghi2018}, linear-quadratic regulator \cite{Won2012}, adaptive control \cite{Hu2007}, variable structure control \cite{HU2005}, and robust control\cite{WU2016}. In particular, adaptive control is applied in~\cite{Yang2005} to a flexible moving beam, and \cite{Zeng1999} utilizes model reference adaptive control to manage a flexible appendage with a rigid central body. Due to the complexity of controlling systems governed by PDEs, a common approach is to discretize the PDE system, reducing it to finite-dimensional ODEs. Then, controllers are developed based on the resulting ODEs; see e.g., \cite{HU2005}. The assumed mode method is popular for this purpose, and extensive control methods have been applied to the ODEs generated from this approach \cite{Turner1984,Hu2008}. However, this model reduction can lead to spillover instability \cite{zhang2011,kar2000}, which occurs when controller is not designed to compensate for higher or unmodeled modes.
 
In recent years, research has shifted towards direct PDE boundary control to address these issues.
Boundary control, in which actuation and sensing occur exclusively through the boundary conditions, applies control effort at the boundary to guide the appendage to its desired position while simultaneously suppressing unwanted vibrations. The authors of \cite{He2015} develop a single control input at the center of the body for a flexible solar panel modeled as two Euler-Bernoulli beams. Meng \cite{meng2016} derives a Lyapunov barrier function that guarantees convergence while maintaining constraints, and  Rad \cite{RAD2018} further advances this work by achieving angle and vibration control without relying on system damping. Additionally, Ma \cite{MA2020} proposes a similar PDE control scheme using the Lyapunov direct method. 

The commonality of these approaches is Lyapunov-based control, which derives control laws from the system model by selecting a Lyapunov candidate function that guarantees stability and convergence. The Lyapunov candidate function typically considers the total system energy, tracking performance, and exponential stability. One drawback of Lyapunov-based control is the intensive derivation process required to formulate the control law, which relies on an accurate pre-derived model.
To overcome these limitations, data-driven control has gained increasing attention for handling complex and nonlinear systems. Unlike traditional methods, data-driven approaches do not require detailed knowledge of the underlying physics. 
One promising technique in this field is \textbf{D}ata-\textbf{e}nabl\textbf{e}d \textbf{p}redictive \textbf{c}ontrol (DeePC), which leverages the fundamental lemma \cite{WILLEMS2005325,Coulson2019}, asserting that for a controllable linear time-invariant (LTI) system, any trajectory of the system can be represented as a linear combination of sequences obtained from a persistently exciting input-output data. DeePC develops non-parametric models from data without the need for explicit system models. This technique has been applied successfully to various applications, including power systems~\cite{huang2021decentralized}, energy management~\cite{SChmitt2024}, quadcopters~\cite{Elokda2021}, soft robots~\cite{Wang2024RAL}, and vehicle platooning \cite{Wang_2023}. The ability to generate control inputs directly from trajectory data makes DeePC particularly suitable for systems with complex, hard-to-model dynamics, such as flexible spacecraft.

This paper makes the following three primary contributions. First, it presents the first application of DeePC to the control of flexible spacecraft that circumvents the need for complicated modeling and controller derivation. Second, it demonstrates the capability of DeePC to track angles and stabilize an infinite number of vibration modes without requiring parametric models or knowledge of system parameters. Third, the proposed method is compared with the well-received Lyapunov-based control method under nominal conditions, uncertainty conditions, and process noise conditions, using a finite element (FE) model.

The remainder of this paper is structured as follows. Section II provides an overview of the modeling of flexible spacecraft. Section III introduces the background of DeePC and its application to the specific problem. Section IV describes the benchmark approaches and compares them with DeePC in numerical simulation. Finally, Section V concludes this study.

\section{System Description} \label{sec_model}

In this section, we present the system description and mathematical modeling of the flexible spacecraft. Since DeePC is a data-driven method, the model will be solely used to generate output data from input data, and the underlying dynamics will not be utilized by DeePC. 

\begin{figure}[!h]
		\centering
		\includegraphics[width=0.9\linewidth]{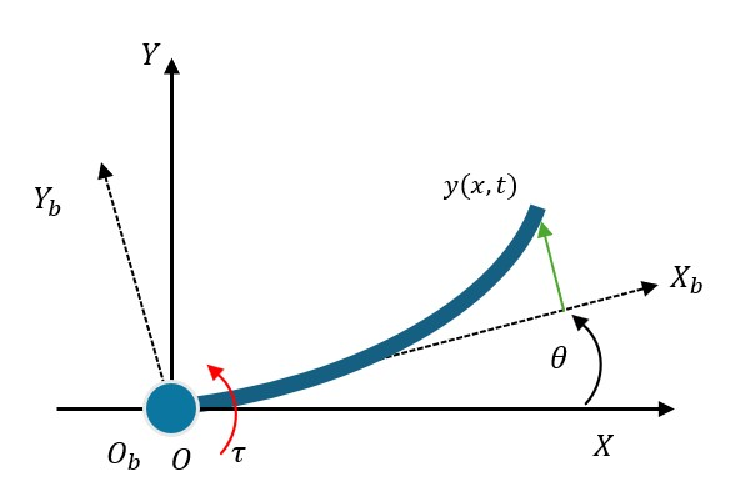}
		\caption{Schematic of the top view of the flexible spacecraft.}\label{fig_diag}
	\end{figure}

The flexible spacecraft of interest is a satellite with a long, flexible antenna. The spacecraft can be modeled as a central hub mass with a flexible cantilever beam, as illustrated in Fig.~\ref{fig_diag}. The flexible beam is modeled as an Euler-Bernoulli beam. The control objective is to guide the flexible beam to a desired angle \(\theta_d\) while suppressing unwanted vibrations. Due to the material's flexibility, the beam deforms during motion, making it challenging to accurately track the angle. In the absence of gravity and friction in space, the flexible beam may vibrate indefinitely.

The coordinate frames are defined as follows: \( XOY \) is the inertial coordinate frame, and \( X_b O_b Y_b \) is the body-fixed coordinate frame. The pitch angle of rotation of the hub is denoted by \( \theta(t) \), and the elastic displacement of the beam at position \(x\) with respect to the body frame is represented by \( y(x, t) \). The position of a point on the appendage in the inertial frame is given by \( z(x, t) = x \theta(t) + y(x, t) \) (Note: the radius of the hub, $r$, is much smaller than the length of the appendage. Therefore, we simplify the rigid-body motion due to base rotation as \(x \theta(t)\) instead of \((x+r) \theta(t)\)). The only available control input, \( \tau \), is the torque applied to the rotating hub. Boundary control is implemented by applying this torque at the hub to achieve the desired motion and vibration suppression.

 The equations of motion for such flexible link systems can be derived using Hamilton's principle~\cite{Cannon1985,Ge1998}. The governing equations for the motion of the system are:

\begin{gather}\label{eq:pde:1}
    J\theta_{tt}(t) - EI y_{xx}(0,t) = \tau(t), \\\label{eq:pde:2}
    \rho y_{tt}(x,t) + EI y_{xxxx}(x,t) + \rho x \theta_{tt}(t) = 0, \\\label{eq:pde:3}
    y(0, t) = 0, \ y_x(0, t) = 0, \\\label{eq:pde:4}
    y_{xx}(L, t) = 0, \ y_{xxx}(L, t) = 0
\end{gather}
The symbols are explained as follows. For the beam, \( E \) is the Young's modulus of the beam, \( I \) is the area moment of inertia of the beam cross section, and \( \rho \) denotes the mass density per unit length of the beam;
at the hub, \( J \) is the moment of inertia of the hub, \( \theta_{tt}(t) \) is the angular acceleration of the hub, $EI y_{xx}(0,t)$ is the bending moment due to beam deflection, and \( \tau(t) \) is the torque applied to the hub;
lastly, the flexible appendage is modeled as a clamped-free beam with boundary conditions given in \eqref{eq:pde:3} and \eqref{eq:pde:4}.

The system comprises both PDE and ODE, presenting the challenge of controlling the hybrid system using only a single control input. As mentioned earlier, discretizing the PDE-ODE system into an ODE system using methods such as the assumed mode method may result in spillover instability. Alternatively, solving the system directly with boundary control involves lengthy derivations and proofs of the Lyapunov function. Furthermore, the model-based control efforts require accurate calibrations on the model parameters such as the beam's Young's modulus $E$ and the area moment of inertia $I$, which can be a costly and time-consuming process. In the next section, we introduce data-driven control, which presents itself as a compelling approach that enables control without requiring extensive derivation of the system dynamics.

\section{DeePC Formulation and Control design} \label{sec_control_objective}

The goal of the spacecraft control problem is to orient the appendage to the desired angle while suppressing deflection. In this section, we will present the control objectives of the flexible spacecraft and explain how the DeePC framework achieves model-free control directly using input-output data. Furthermore, we will introduce how dimension reduction is incorporated into the DeePC approach.

\subsection{ Data-Enabled Predictive Control} \label{sub_deepc}

A brief overview of DeePC~\cite{Coulson2019}  and important preliminaries are now introduced.
Consider the parametric model of a linear time-invariant (LTI) system represented as follows:
\begin{equation}\label{eq:parametric_description}
\begin{aligned}
    x_{k+1} &= A x_k + B u_k, \\
    y_k &= C x_k + D u_k,
\end{aligned}
\end{equation}
where $A \in \mathbb{R}^{n \times n}$, $B \in \mathbb{R}^{n \times m}$, $C \in \mathbb{R}^{p \times n}$, $D \in \mathbb{R}^{p \times m}$ are system matrices; $x_k \in \mathbb{R}^m$ is the state vector, $u_k \in \mathbb{R}^m$ is the input vector, and $y_k \in \mathbb{R}^p$ is the output vector. Despite this linear system representation, it has a wide range of applications, as a nonlinear system can be linearized around the operating point to capture its local dynamics. For the flexible spacecraft, the system states include all underlying dynamics related to the Euler-Bernoulli beam, such as the beam deflection and its derivatives. The system output is the tip-end deflection, and the input is the control torque.

To represent the system described in~\eqref{eq:parametric_description} as a non-parametric model, the DeePC algorithm builds its foundation on Willems' fundamental lemma~\cite{WILLEMS2005325}. This approach can replicate the system's behavior, where the dynamics can be characterized solely based on input and output data. A key requirement for this is that the input data must be \textit{persistently exciting}, which is defined below:

\begin{definition}
Given a signal $\omega_{\left[0, T-1\right]}\!:=\!\begin{bmatrix}
	  \omega^{\tr}(0), \omega^{\tr}(1), \cdots, \omega^{\tr}(T-1)
    \end{bmatrix}^{\tr}$, the corresponding Hankel matrix is of depth $L$ and length $T$ (where $L, T\in \mathbb{Z}$ and $L\le T$) and defined as:
    \begin{equation}
    \mathcal{H}_{L}(\omega_{\left[0, T-1\right]}):=
	\begin{bmatrix}
		\omega(0) & \omega(1) & \cdots & \omega(T-L) \\
		\omega(1) & \omega(2) & \cdots & \omega(T-L+1)\\
		\vdots & \vdots & \ddots & \vdots \\
		\omega(L-1) & \omega(L) & \cdots & \omega(T-1)
	\end{bmatrix}.
    \end{equation}
	The sequence $\omega_{[0,T-1]}$ is said to be \textit{persistently exciting of order $L$} 
	if $\mathcal{H}_{L}(\omega_{[0,T-1]})$ maintains full row rank. 
\end{definition}

\begin{lemma}[Fundamental Lemma \cite{WILLEMS2005325}] \label{lemma 1}
    Let a sequence $(u_{[0,T-1]}^{\mathrm{d}}$ $y_{[0,T-1]}^{\mathrm{d}})$ be an input/output trajectory of controllable LTI system~\eqref{eq:parametric_description}, where  $u_{[0,T-1]}^{\mathrm{d}}$ is persistently exciting of order $n + L$. Then, any length-$L$ sequence $(u_{[0,L-1]}, y_{[0,L-1]})$ is an input/output trajectory of \eqref{eq:parametric_description} if and only if there exists real vector $g \in \mathbb{R}^{(T - L + 1)}$ such that 
    \begin{equation} \label{eq:fundamental-lemma}
	\begin{bmatrix}
		u_{[0,L-1]} \\ y_{[0,L-1]}
	\end{bmatrix} = \begin{bmatrix}
		\mathcal{H}_{L}(u^{\mathrm{d}}_{[0,T-1]}) \\
		\mathcal{H}_{L}(y^{\mathrm{d}}_{[0,T-1]})
	\end{bmatrix} g.
	\end{equation}   
\end{lemma}

Based on Lemma~\ref{lemma 1}, the Hankel matrix that is capable of representing the system can be constructed as the Hankel matrix of the input, $\mathcal{H}_L(u_{[0, T-1]}^{\mathrm{d}})$, concatenated with the Hankel matrix of the output, $\mathcal{H}_L(y_{[0, T-1]}^{\mathrm{d}})$:

\begin{equation}\label{eq:non_parametric_description}
    \resizebox{\columnwidth}{!}{
        $\begin{aligned}
            \begin{bmatrix}
                \mathcal{H}_L(u^{\mathrm{d}}_{[0, T-1]}) \\
                \hline
                \mathcal{H}_L(y^{\mathrm{d}}_{[0, T-1]})
            \end{bmatrix}
            &:= 
            \begin{bmatrix}
                u^{\mathrm{d}}(0) & u^{\mathrm{d}}(1) & \dots & u^{\mathrm{d}}(T - L) \\
                \vdots & \vdots & \ddots & \vdots \\
                u^{\mathrm{d}}(L - 1) & u^{\mathrm{d}}(L) & \dots & u^{\mathrm{d}}(T - 1) \\
                \hline
                y^{\mathrm{d}}(0) & y^{\mathrm{d}}(1) & \dots & y^{\mathrm{d}}(T - L) \\
                \vdots & \vdots & \ddots & \vdots \\
                y^{\mathrm{d}}(L - 1) & y^{\mathrm{d}}(L) & \dots & y^{\mathrm{d}}(T - 1) \\
            \end{bmatrix}.
        \end{aligned}$
    }
\end{equation}
To collect data to generate the matrix to represent~\eqref{eq:parametric_description}, we need the input matrix $u_{\left[0, T-1\right]}^{\mathrm{d}}$ and the output matrix $y_{\left[0, T-1\right]}^{\mathrm{d}}$, whose sequences of length $T$ are shown below:
\begin{equation}
    \begin{aligned}
    u_{\left[0, T-1\right]}^{\mathrm{d}}&:=\begin{bmatrix}
	u^{\mathrm{d}}(0)^{\tr}, u^{\mathrm{d}}(1)^{\tr}, \cdots, u^{\mathrm{d}}(T-1)^{\tr}
    \end{bmatrix}^{\tr},
   \\
   y_{\left[0, T-1\right]}^{\mathrm{d}}&:=\begin{bmatrix}
	y^{\mathrm{d}}(0)^{\tr}, y^{\mathrm{d}}(1)^{\tr}, \cdots, y^{\mathrm{d}}(T-1)^{\tr}
    \end{bmatrix}^{\tr}.
    \end{aligned}
\end{equation}
 For the DeePC formulation, we split the matrix into past data section and a future data section. We define $T_{\mathrm{ini}}, N \in \mathbb{Z}$ as the length of past and future data. The depth is $L$ and is obtained as $T_{\mathrm{ini}} + N=L$. The above Hankel matrices $\mathcal{H}_L(u_{[0, T-1]}^{\mathrm{d}})$ and $\mathcal{H}_L(y_{[0, T-1]}^{\mathrm{d}})$ can be then split into two parts:
 
\begin{equation}\label{eq:UP_Uf}
\begin{aligned}
    \begin{bmatrix}
        U_{\mathrm{p}} \\ U_{\mathrm{f}}
    \end{bmatrix}
    := \mathcal{H}_L(u_{[0, T-1]}^{\mathrm{d}}), \quad
    \begin{bmatrix}
        Y_{\mathrm{p}} \\ Y_{\mathrm{f}}
    \end{bmatrix}
    = \mathcal{H}_L(y_{[0, T-1]}^{\mathrm{d}}),
\end{aligned}
\end{equation}
where \( U_{\mathrm{p}} \) consists of the first \( T_{\mathrm{ini}} \) block rows of \( \mathcal{H}_L(u_{[0, T-1]}^{\mathrm{d}}) \), and \( U_{\mathrm{f}} \) consists of the last \( N \) block rows of \( \mathcal{H}_L(u_{[0, T-1]}^{\mathrm{d}}) \). Similarly, the same applies to \( Y_{\mathrm{p}} \) and \( Y_{\mathrm{f}} \).

 Governed by Willems' fundamental lemma, we aim to predict the future control sequence of length \( N \) based on the past control sequence of length \( T_{\mathrm{ini}} \). We define \( u_{\mathrm{ini}} = u_{[k-T_{\mathrm{ini}}, k-1]} \) as the past control sequence and \( u = u_{[k, k+N-1]} \) as the predicted future control sequence. The same logic applies to \( y_{\mathrm{ini}} \) and \( y \). This optimization is formulated below and solved at each time step:

\begin{equation}\label{eq:deepc_formulation}
    \begin{aligned}
        \min_{g, u, y} \quad & \|y - y_r\|_{Q}^2 + \|u\|_{R}^2
        \\
        \text{subject to} \quad & \begin{bmatrix} U_{\mathrm{p}} \\ U_{\mathrm{f}} \\ Y_{\mathrm{p}} \\ Y_{\mathrm{f}} \end{bmatrix} g = \begin{bmatrix} u_{\mathrm{ini}} \\ u \\ y_{\text{ini}} \\ y \end{bmatrix}, u \in \mathcal{U}, y \in \mathcal{Y},
    \end{aligned}
\end{equation}
where \( y_{r} = \begin{bmatrix}
	y_{r,k}^\top, y_{r,k+1}^\top, \dots, y_{r,k+N-1}^\top
\end{bmatrix}^\top \) is a desired trajectory. $\|y-y_{r}\|_{Q}^2:=(y-y_{r})^{\top} Q (y-y_{r})$. $\|u\|_{R}^2:=u^{\top} R u$.
$Q, R$ are weighting matrices and $\mathcal{U}$, $\mathcal{Y}$ represent the input and output constraints.

The formulation in~\eqref{eq:deepc_formulation} represents the formulation of DeePC in a deterministic LTI system. In real systems, the presence of output measurement noise or system nonlinearities requires modifications and extensions to the DeePC algorithm ~\cite{Coulson2019,Berberich2020}. A slack variable is added to avoid constraint violations caused by measurement noise. The regularization term is added to enhance robustness. Consequently, the regularized DeePC formulation is expressed as follows:

\begin{equation}\label{eq:deepc_regu}
    \begin{aligned}
        \min_{g, u, y, \sigma_y} \quad & \|y - y_r\|_{Q}^2 + \|u\|_{R}^2 + \lambda_y \|\sigma_y\|_2^2 + \lambda_g\|g\|_2^2 \\
        \text{subject to} \quad & \begin{bmatrix} U_{\mathrm{p}} \\ U_{\mathrm{f}} \\ Y_{\mathrm{p}} \\ Y_{\mathrm{f}} \end{bmatrix} g = \begin{bmatrix} u_{\mathrm{ini}} \\ u \\ y_{\text{ini}} \\ y \end{bmatrix} + \begin{bmatrix} 0 \\ 0 \\ \sigma_y \\ 0 \end{bmatrix}, u \in \mathcal{U}, y \in \mathcal{Y}.
    \end{aligned}
\end{equation}
In~\eqref{eq:deepc_regu}, the equality constraint can be relaxed with a slack variable $\sigma_{y}$, which is subject to a weighted quadratic norm penalty function. The weight coefficient $\lambda_y>0$ should be purposefully selected to be sufficiently large to ensure that the cost associated with $\sigma_y \neq 0$ is only incurred when the equality constraint is no longer valid~\cite{Coulson2019,Elokda2021}.
A quadratic norm penalty is applied to $g$ with a weight coefficient $\lambda_{g} > 0$, this term improves robustness when encountering disturbances and nonlinearity data~\cite{Huang2021IFAC,SChmitt2024}.

DeePC handles the nonlinearity and non-deterministic characteristics by introducing additional terms, as shown in the formulation in~\eqref{eq:deepc_regu}. A flexible spacecraft, being a nonlinear system operating in an unpredictable environment, presents itself as a good candidate for such an approach. Algorithm~\ref{algo:deepc} outlines the application of DeePC to a flexible spacecraft. To start the algorithm, we first collect control sequence $u_{\mathrm{ini}}$ and output sequence $y_{\mathrm{ini}}$ of length $T_{\mathrm{ini}}$.
The control sequence is initialized as all zeros, and the output is recorded based on the simulation output of the end-tip angle.
The DeePC starts after $t > T_{\mathrm{ini}}$. After each step, it generates the optimal $g$, the optimal $g$ is then used to optimal predictions for $u$ and $ y$ based on \eqref{eq:deepc_regu}, the process continues until the end of the simulation is reached.

\begin{algorithm} 
  \caption{DeePC Algorithm for Flexible Spacecraft} \label{algo:deepc}
  \begin{algorithmic}[1] %
    \State \textbf{Input:} Total time step $k_{\mathrm{end}}$, pre-collected control input (applied torque) sequence $u^{\mathrm{d}}_{[0, T-1]}$ and spacecraft angle output sequence $y^{\mathrm{d}}_{[0, T-1]}$.
    \State Construct Hankel matrices $U_{\mathrm{p}}$, $U_{\mathrm{f}}$, $Y_{\mathrm{p}}$, $Y_{\mathrm{f}}$.
    \State For $k<T_{\text{ini}}$, initialize $u_{\text{ini}}$ with 0 and $y_{\text{ini}}$ with corresponding angle.
    \While {$T_{\text{ini}}\le k \le k_{\mathrm{end}}$} \label{algo:line:startwhile}
      \State DeePC optimization \eqref{eq:deepc_regu} solves for $g$ and provide optimal control $u=U_{\mathrm{f}}g$.
      \State  Apply the first step optimal control torque $u(1)$.
      \State  Obtain the new angle and update $u_{\mathrm{ini}}$ and $y_{\mathrm{ini}}$ to the $T_{\mathrm{ini}}$ most recent input/output measurements.
      \State  Set $k$ to $k+1$.
    \EndWhile \label{algo:line:endwhile}
  \end{algorithmic}
\end{algorithm}

\subsection{Dimension Reduction for DeePC} \label{sub_minimum}
For $u^{\mathrm{d}}_{[0,T-1]}$ to satisfy the requirement of persistent excitation, column number must be no less than its row number. As a result, the length $T$ must satisfy $T - (n + L) + 1 \geq m(n+L)$, i.e., $T \geq (m+1)(n+L) -1$. Thus, the dimension of $g$ in \eqref{eq:deepc_regu} is lower bounded as
\begin{equation} \label{eq:g-dimension}
	T-L+1 \geq mL + (m+1)n.
\end{equation}

As we can see from \eqref{eq:g-dimension}, if the length of \( T \) is large, it could cause large dimension variable \( g \) in \eqref{eq:deepc_regu}. In order to collect sufficient data for persistent excitation, it needs to ensure \( T \geq (m+1)(n+L) - 1 \). The issue arises with large \( T \) causes high dimension for optimization problem in \eqref{eq:deepc_regu}, making it computationally slow to solve.

To address issues caused by large dimensions, we apply a singular value decomposition (SVD)-based approach~\cite{ZhangCSL2023} to reduce the dimensionality of the optimization variables in DeePC, thereby improving computational efficiency. The goal of using SVD-based approach is to obtain the most important features from the original Hankel matrix, reducing computational cost while maintaining performance.

The SVD decomposition is applied to the original Hankel matrices in \eqref{eq:non_parametric_description}, which consists of  $u_{[0, T-1]}^{\mathrm{d}}$ and $y_{[0, T-1]}^{\mathrm{d}}$. The results are expressed in the following form:

\begin{equation}\label{eq:SVD_decomposition}
\begin{bmatrix}
    \mathcal{H}_{L}(u^{\mathrm{d}}_{[0, T-1]}) \\
    \mathcal{H}_{L}(y^{\mathrm{d}}_{[0, T-1]})
\end{bmatrix}
=
\underbrace{\begin{bmatrix}
	W_1 & W_2
\end{bmatrix}}_{W} \underbrace{\begin{bmatrix}
	\Sigma_1 & 0 \\ 0 & \Sigma_2
\end{bmatrix}}_{\Sigma} \underbrace{\begin{bmatrix}
	V_1 & V_2
\end{bmatrix}^\tr}_{V^{\tr}},
\end{equation}
where $W\in \mathbb{R}^{q_{1}\times q_{1}}$ and $V\in \mathbb{R}^{q_{2}\times q_{2}}$ are the orthogonal matrices consisting of the left and right singular vectors, respectively. Here, \( q_{1} = (m+p)L \) and \( q_{2} = T-L+1 \). $\Sigma\in \mathbb{R}^{q_1\times q_{2}}$ is the rectangular diagonal matrix with non-negative singular values arranged in decreasing order along its diagonal.
We aim to define reduced-dimension Hankel matrix that retains the essential information as the original. We construct $\bar{\mathcal{H}}_L \in \mathbb{R}^{q_1 \times r}$ as:
\begin{equation} \label{eq:newHL}
	\bar{\mathcal{H}}_{L} = \mathcal{H}_{L}V_{1} = W_1 \Sigma_{1}.
\end{equation}
$\Sigma_1 \in \mathbb{R}^{r \times r}$ contains the first $r$ non-zero singular values in decreasing order and $r\le \min\{q_1, q_2\}$. $\Sigma_2$ corresponds to the zero sections of the singular values.  $W_{1}$ and  $V_{1}$ are matched with correct dimension for $\Sigma_1$  and the same applies to $W_{2}$ and  $V_{2}$ for $\Sigma_2$. With $\bar{\mathcal{H}}_L$ defined, we construct the reduced version of~\eqref{eq:deepc_regu}, as shown below:
\begin{equation}\label{eq:deepc_svd}
\begin{aligned}
    \min_{\bar{g}, u, y, \sigma_y} \quad & \|y - y_r\|_{Q}^2 + \|u\|_{R}^2 + \lambda_y \|\sigma_y\|_2^2 + \lambda_g\|\bar{g}\|_2^2 \\
    \text{subject to} \quad & \bar{\mathcal{H}}_{L} \bar{g} = \begin{bmatrix} u_{\mathrm{ini}} \\ u \\ y_{\text{ini}} \\ y \end{bmatrix} + \begin{bmatrix} 0 \\ 0 \\ \sigma_y \\ 0 \end{bmatrix}, u \in \mathcal{U}, y \in \mathcal{Y}.
\end{aligned}
\end{equation}

For an LTI system, the new matrix $\bar{\mathcal{H}}_L$ preserves the same range space as the original Hankel matrix, since $\Sigma_1$ contains all non-zero singular values $r$ and it is equal to the rank of the matrix~\eqref{eq:non_parametric_description}). In real practice, the system will almost always appear full rank due to process and measurement noise. In such cases, it is necessary to determine an appropriate cutoff for $r$. We investigate this cutoff based on the singular value distribution. By applying the method in~\cite{ZhangCSL2023}, a turning point is identified in the singular value distribution, providing a metric to separate key features from those that can be neglected.

With the SVD method applied, the original optimization variable $g$ in~\eqref{eq:deepc_regu}, of size $T - L + 1$, is reduced to the optimization variable $\bar{g}$ of dimension $r$. The more compact data matrix $\bar{\mathcal{H}}_L$ speeds up the computational process for DeePC and enables it to tackle more complex systems.
	
\section{Simulations} \label{sec_experiments}
This section presents an experimental study of DeePC's performance on the flexible spacecraft system, with results compared to a well-established Lyapunov-based method. To demonstrate the effectiveness of the proposed control methods, numerical simulations are conducted using FE method. The spatial and temporal discretization steps are set to 0.25 m and 0.05 s, respectively. The system parameters are as follows: $EI = 120 \, \mathrm{kg \cdot m^3/s^2}$, $J = 400 \, \mathrm{kg \cdot m^2}$, $\rho = 20 \, \mathrm{kg/m}$, and $L = 5 \, \mathrm{m}$.  More details on the FE model are provided in Appendix \ref{sec_appendix_fe}.

\begin{figure}[!h]
		\centering
		\includegraphics[width=1\linewidth]{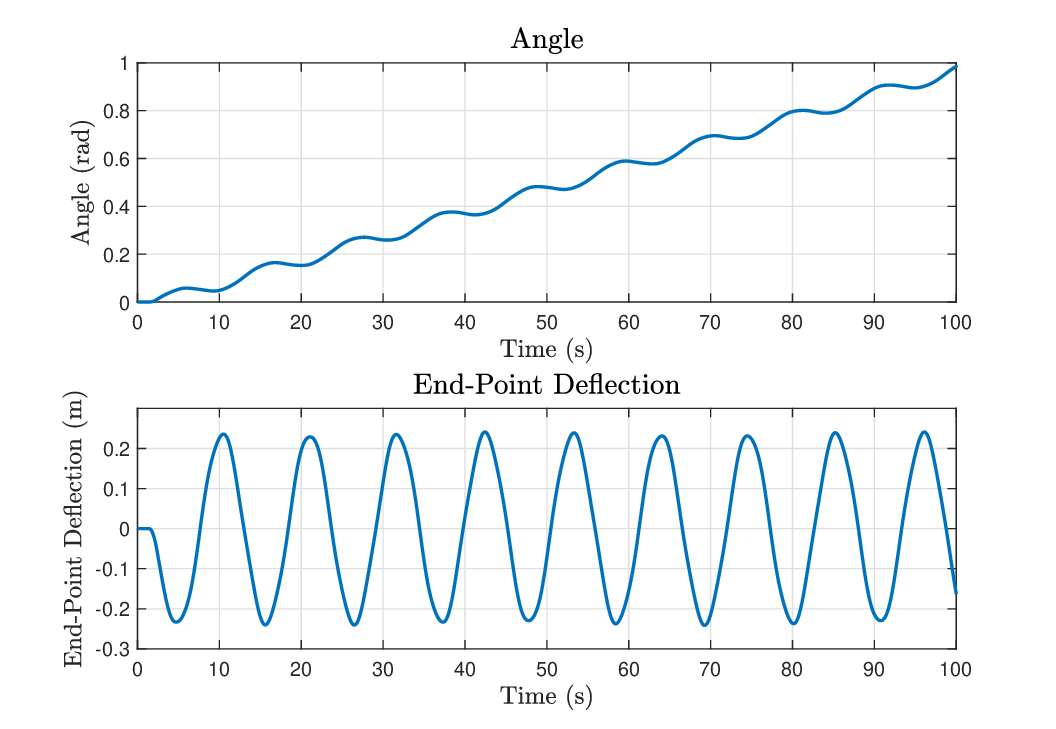}
		\caption{The angle and end-point deflection of the flexible appendage under no control. }\label{fig_free_angle}
	\end{figure}

\begin{figure}[!h]
		\centering
		\includegraphics[width=1\linewidth]{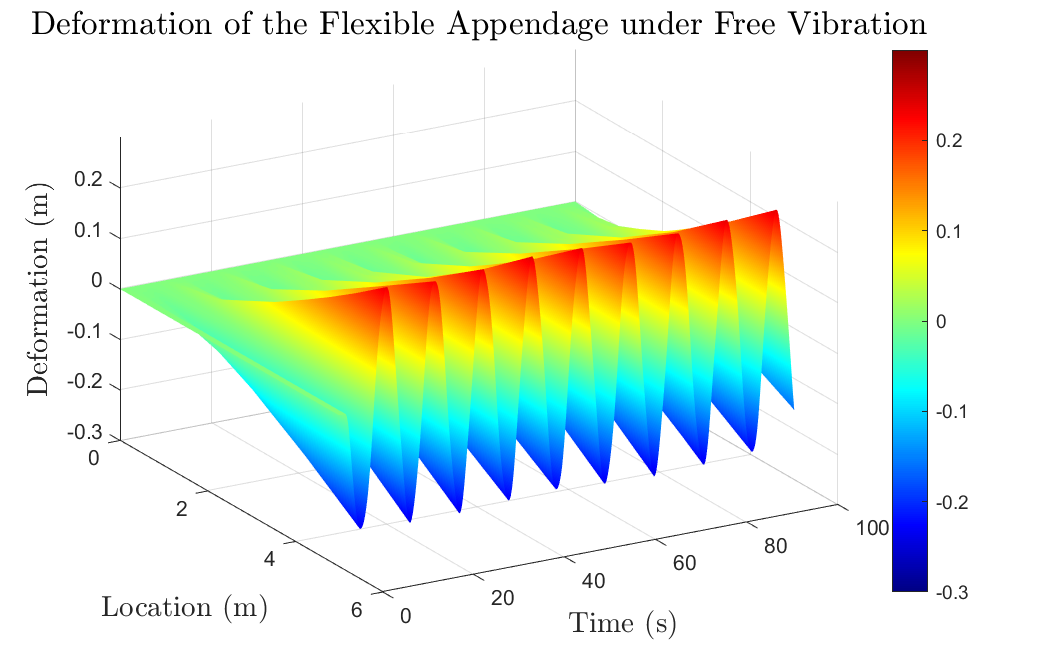}
		\caption{The deformation of the flexible appendage under no control. }\label{fig_free_vibration}
	\end{figure}

We first examine the system behavior without control. Due to the lack of damping in the space environment, if the spacecraft is subjected to an impulse torque without subsequent control effort, the angular motion will continue indefinitely. As shown in Fig.~\ref{fig_free_angle}, when the system is subjected to an impulse of $5 \, \mathrm{Nm}$, it continues to oscillate. Additionally, in the absence of friction and gravity, the appendage vibrates indefinitely. The deformation over time is illustrated in Fig.~\ref{fig_free_vibration}.

        \subsection{Setup and Data Collection} \label{sub_setup}

Non-parametric representations are obtained through offline data collection, which involves gathering trajectories that represent the system's behavior. To construct such a matrix, a PD controller is used during data collection to track several desired angles. It is important to note that, as demonstrated in a previous study \cite{MA2020}, a simple PD controller fails to achieve stability and convergence, resulting in the flexible beam vibrating around the desired location. The control inputs and corresponding outputs are recorded. The sufficiently rich dataset captures the underlying dynamics of the complex system. Subsequently, this data is organized into a Hankel matrix using the method described in Sec.~\ref{sub_deepc}.

    	\begin{figure*}[!h]
		\centering
		\includegraphics[width=1\linewidth]{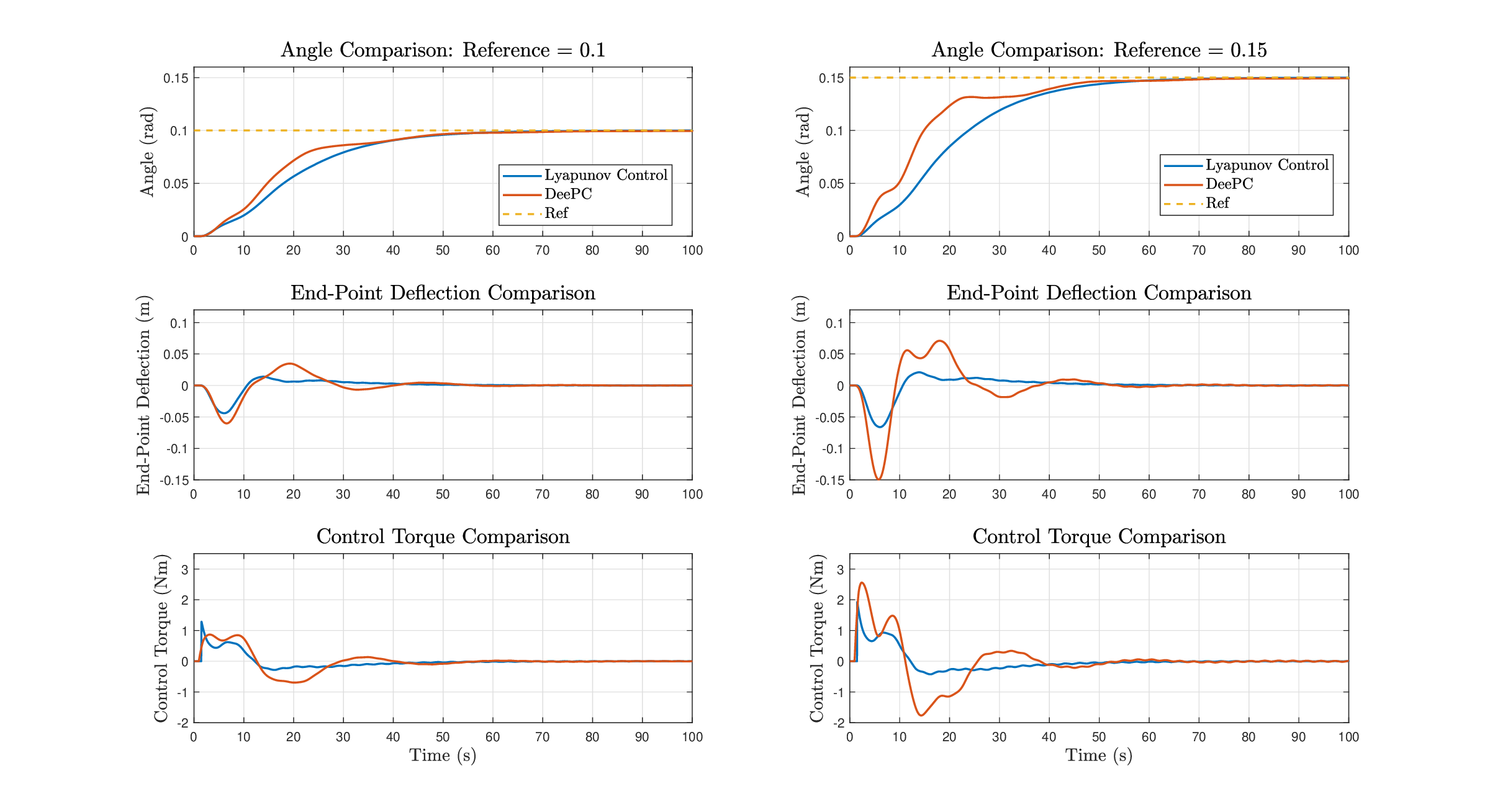}
		\caption{The performance comparison between DeePC and Lyapunov-based control under nominal case. }\label{fig_perf1}
	\end{figure*}

        \subsection{Benchmark Approach: Lyapunov-based Control Tracking Results} \label{sub_model_based}

In this section, we present a Lyapunov-based control method. A control law that achieves asymptotic stability of \( e(t) \to 0 \) and stability of \( \|y(\cdot, t)\|_{L_2} \) can be straightforwardly designed using Lyapunov functionals that represent the total energy of the system \cite{Yigit1994, Ge1998}. A control that achieves asymptotic stability of both is introduced by Rad in \cite{RAD2018Acta}. We modified the approach from Rad and proposed our own Lyapunov candidate function.

Consider the following Lyapunov functional candidate :

\begin{equation}\label{eq:ly_candidate}
V(t) = E_a(t) + \frac{EIa_1}{2} e^2(t) + \frac{1}{2} \Delta^2(t) + a_3 V_1(t) + a_4 V_2(t),
\end{equation}
where \( E_a \) denotes the sum of kinetic and potential energy of the appendage:
\begin{equation}
E_a(t) = \frac{\rho}{2} \int_0^L z_t^2(x,t) \, dx + \frac{EI}{2} \int_0^L y_{xx}^2(x,t) \, dx,
\end{equation}
with
\begin{equation}
\Delta(t) = J \theta_t(t) + a_2 \left( -y_{xx}(0,t) + a_1 e(t) \right),
\end{equation}

\begin{equation}
V_1(t) = \rho \int_0^L (x - L) z_t(x,t) y_x(x,t) \, dx,
\end{equation}

\begin{equation}
V_2(t) = \rho \int_0^L z_t(x,t) \left( x e(t) + y(x,t) \right) \, dx.
\end{equation}
$\Delta$, $V_1$, and $V_2$ are selected to ensure exponential stability. The details and proof of stability of the Lyapunov candidate function are provided in the Appendix \ref{sec_appendix_stability}. The following control law is considered:

\begin{align}
\tau(t) &= \left( - EI y_{xx}(0,t) + a_2 y_{xxt}(0,t) - a_1 a_2 \theta_t(t) \right) \notag \\
&\quad - \frac{EI}{a_2} \theta_t(t) - k_1 \Delta(t) \notag \\
&\quad + k_2 \left( J \theta_t(t) + a_2 y_{xx}(0,t) - a_1 a_2 e(t) \right).
\end{align}

The parameters must satisfy the following conditions:

\begin{equation}
a_1, a_2, a_3, a_4, k_1, k_2, \delta > 0
\end{equation}

\begin{equation}
1 - L a_3 - a_4 > 0,
\end{equation}

\begin{equation}
\frac{EI}{2} - \frac{2 \rho L^3 a_3}{\pi^2} - \frac{8 \rho L^4 a_4}{\pi^4} > 0,
\end{equation}

\begin{equation}
\frac{EI a_1}{2} - \frac{2 \rho L^3 a_4}{\pi^2} > 0,
\end{equation}

\begin{equation}
\frac{J EI}{a_2} - k_2 J^2 - \frac{\rho L^2 a_3}{2 \delta} - \frac{2 \rho L^3 a_4}{3} > 0,
\end{equation}

\begin{equation}
(1 - L \delta) a_3 - 4 a_4 > 0,
\end{equation}

\begin{equation}
k_2 a_2^2 - \frac{LEI a_3}{2} \geq 0.
\end{equation}

The control parameters satisfying these requirements are selected as:
\( a_1 = 0.0428 \),
\( a_2 = 3000 \),
\( k_1 = 0.1 \),
and \( k_2 = 2.1 \times 10^{-10} \).

\subsection{Proposed Approach: DeePC Tracking Results} \label{sub_DeePC_results}

The parameter selections for DeePC are:
\( T_{\text{ini}} = 20 \),
\( N = 20 \),
\( Q = 1000 \),
\( R = 0.25 \times 10^{-3} \),
\( \lambda_g = 1000 \),
and \( \lambda_y = 300000 \). As described in Sec.~\ref{sub_minimum}, the original Hankel matrix has a size of \( 80 \times 3961 \). After applying the dimension reduction techniques described in~\eqref{eq:newHL}, the reduced Hankel matrix is of size of \( 80 \times 80 \).

 \subsubsection{Nominal Case} \label{nominal_case}
\begin{figure}[!h]
		\centering
		\includegraphics[width=1\linewidth]{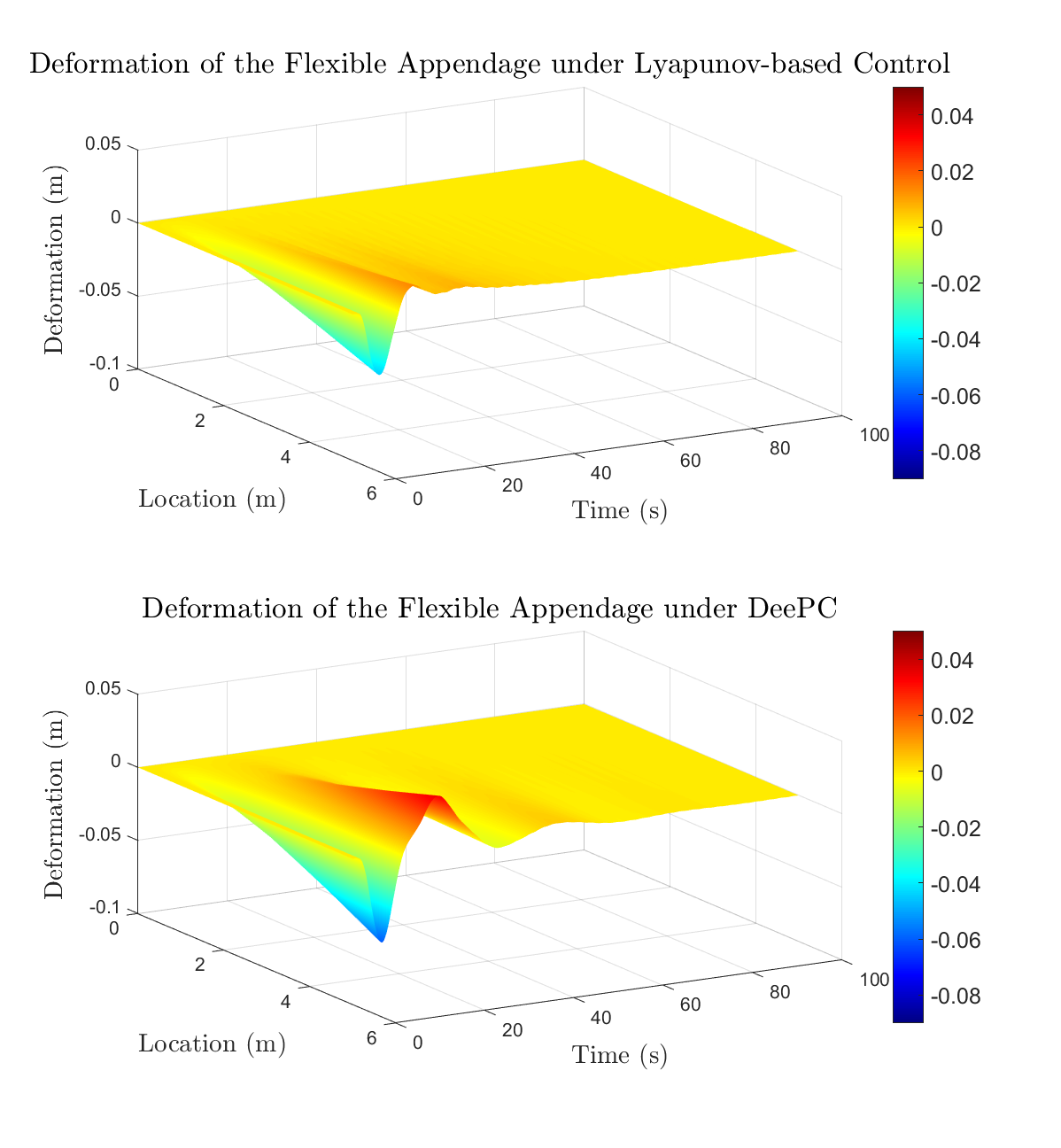}
		\caption{The deformation of the flexible spacecraft with Lyapunov-based control vs DeePC control under nominal conditions. }\label{fig_def1}
	\end{figure}

In the first scenario, we assume the exact model parameters of the spacecraft are perfectly known to the Lyapunov method (note that this is not needed for DeePC). Using the non-parametric formulation obtained from Sec.~\ref{sub_deepc}, we perform DeePC control based on the approach described in \eqref{eq:deepc_svd}. For comparison, we also implement a Lyapunov-based control method, leveraging the exact parameters, and evaluate its performance in angle tracking. The results of both methods are compared, as illustrated in Fig.~\ref{fig_perf1}. DeePC achieves a similar convergence speed to Lyapunov-based control but requires a higher control input. However, DeePC yields a lower overall cost based on the defined cost function. The result is expected, as the initial larger deviation from the desired position results in a higher cost, and predictive control effectively addresses this issue by bringing the angle to the desired value more quickly. The deflection convergence is illustrated in Fig.~\ref{fig_def1}, where it can be observed that both approaches suppress deflection and achieve convergence.

\subsubsection{Model Uncertainty and Process Noise} \label{model uncertainty and process noise}

\begin{figure}[!h]
		\centering
		\includegraphics[width=1\linewidth]{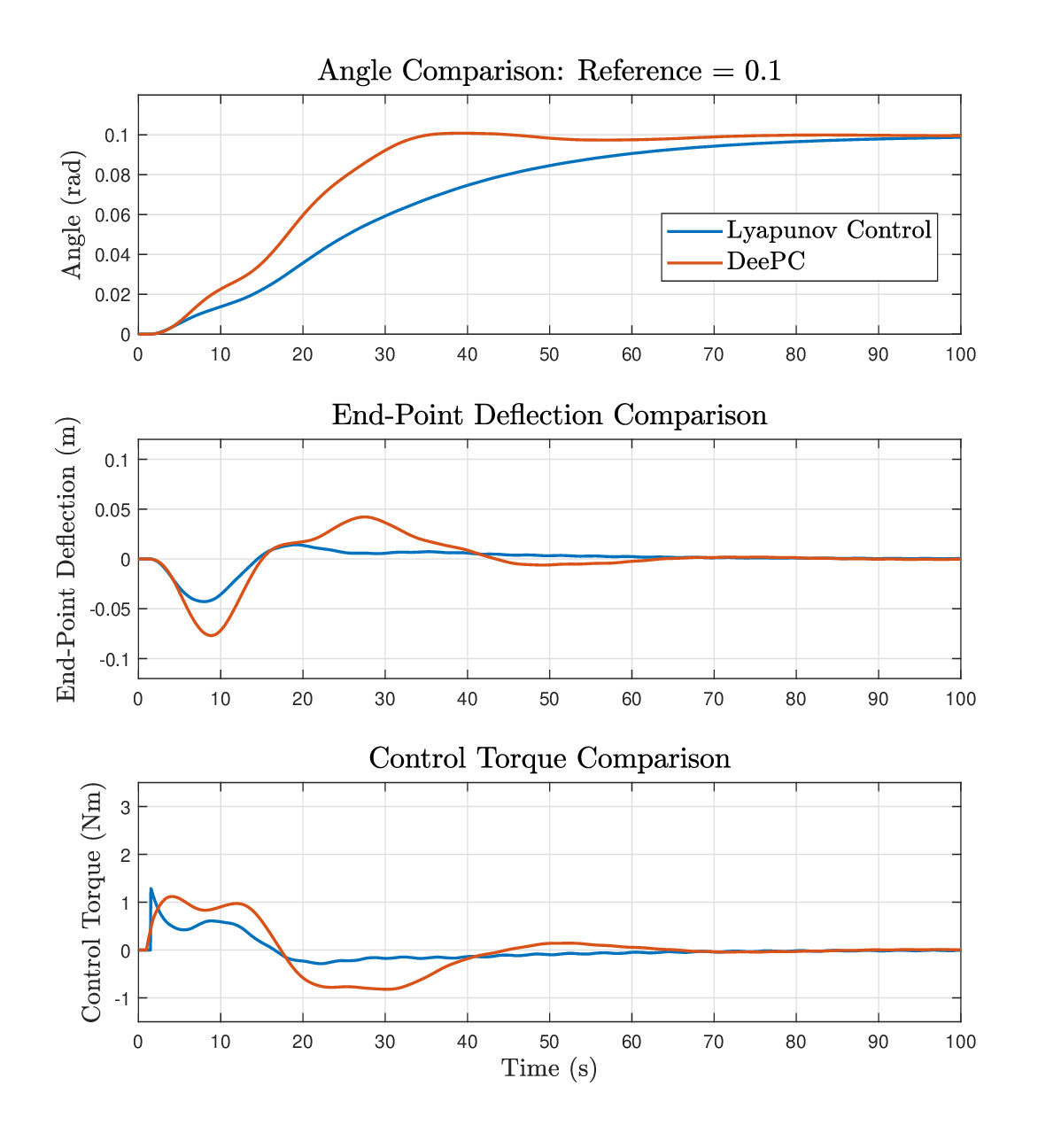}
		\caption{The performance comparison between Lyapunov-based control and DeePC under model uncertainty. }\label{fig_perf2}
	\end{figure}

One advantage of the DeePC formulation is that it does not rely on a pre-derived mathematical model; instead, it utilizes input and output data to create a non-parametric representation. When model uncertainty is introduced, where the actual weight and material density are twice those assumed in the model, a performance comparison of the two methods is shown in Fig.~\ref{fig_perf2}. While the system still converges, the rise time increases for Lyapunov-based control. In contrast, DeePC maintains its fast rise time with a lower cost despite the uncertainty in the model. The robustness of DeePC arises from the term \( \lambda_g \|g\|_2^2 \), which is implemented to render the cost function quadratic. This formulation leads to robust and optimal solutions with respect to bounded disturbances in the input/output data~\cite{Huang2021IFAC}.

\begin{figure}[!h]
		\centering
		\includegraphics[width=1\linewidth]{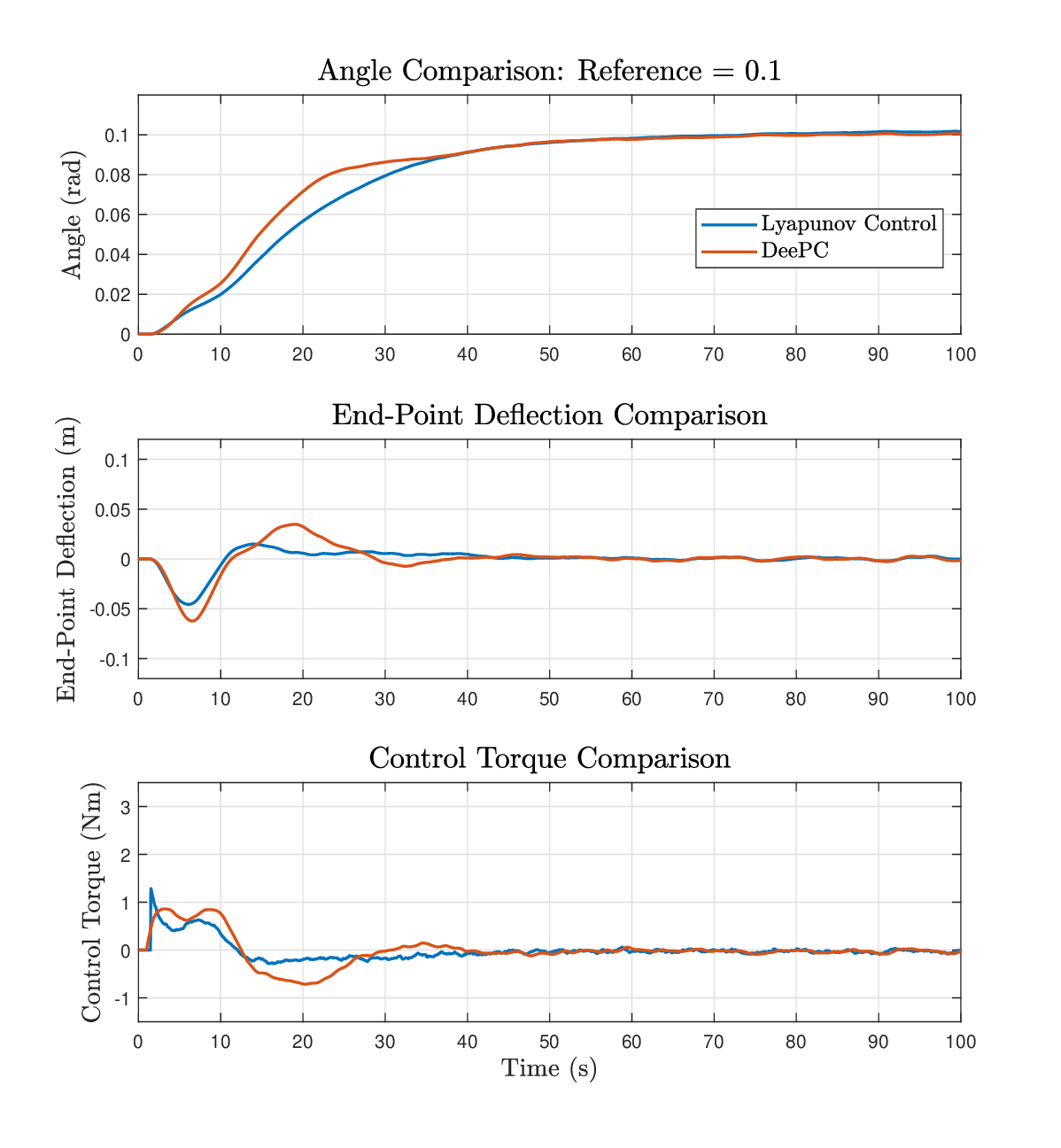}
		\caption{The performance comparision between Lyapunov-based control and DeePC under process noise. }\label{fig_perf3}
	\end{figure}

\begin{figure}[!h]
		\centering
		\includegraphics[width=1\linewidth]{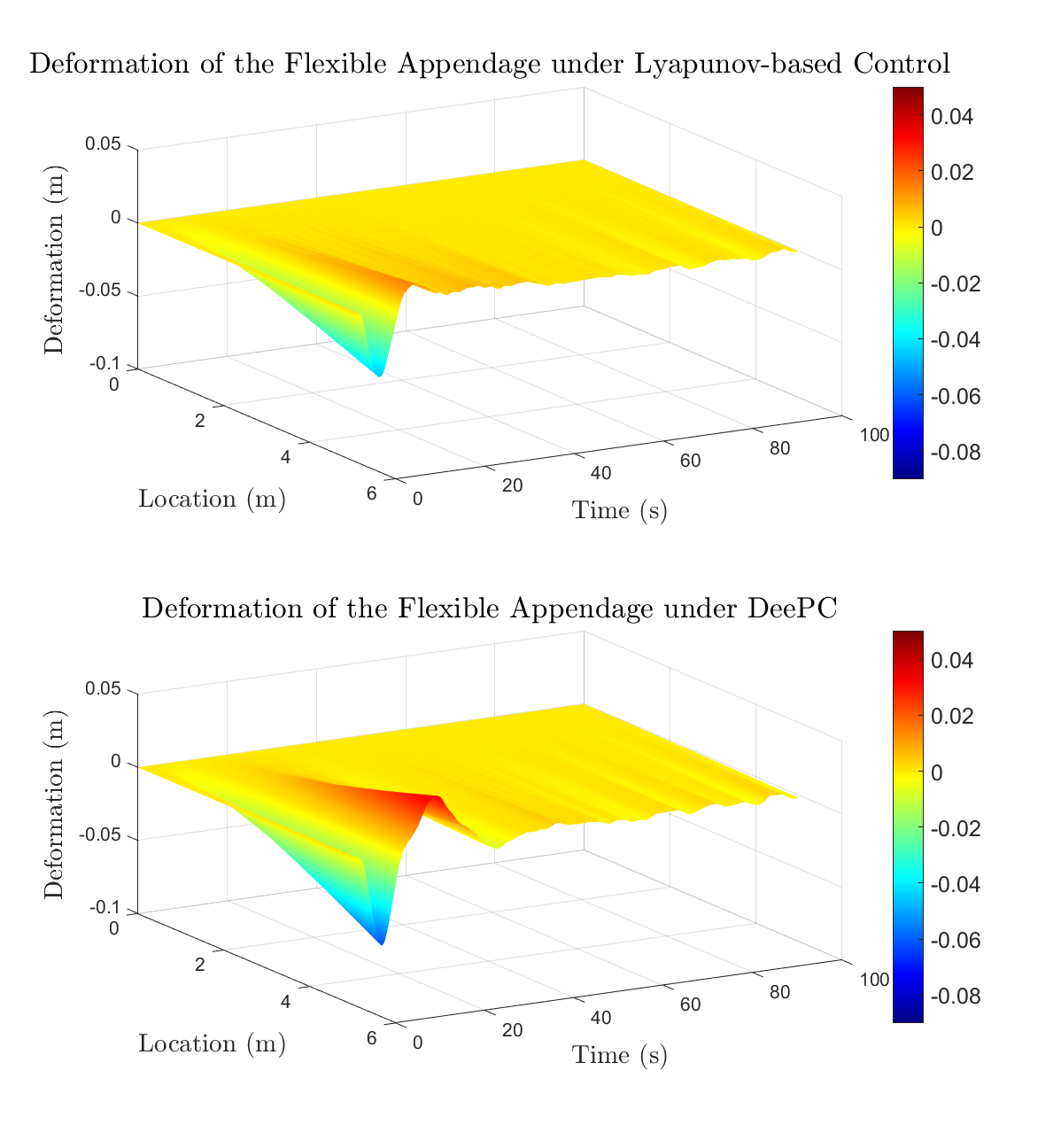}
		\caption{The deformation of the flexible spacecraft with Lyapunov-based control vs DeePC control under process noise. }\label{fig_def3}
	\end{figure}

In an unpredictable environment like space, flexible structures are frequently subjected to disturbances, such as cosmic winds, which can significantly impact their behavior. These disturbances are modeled as process noise in our study. To evaluate the robustness of the control methods under such challenging conditions, we compare the performance of DeePC and the Lyapunov-based control method. Despite the presence of process noise, both methods successfully maintain accurate angle tracking, demonstrating their effectiveness in dealing with external perturbations. The comparative results are depicted in Fig.~\ref{fig_perf3} and Fig.~\ref{fig_def3}, highlighting the ability of both approaches to ensure stable performance and reliable tracking in a noisy environment.

     	\begin{figure}[!h]
		\centering
		\includegraphics[width=1\linewidth]{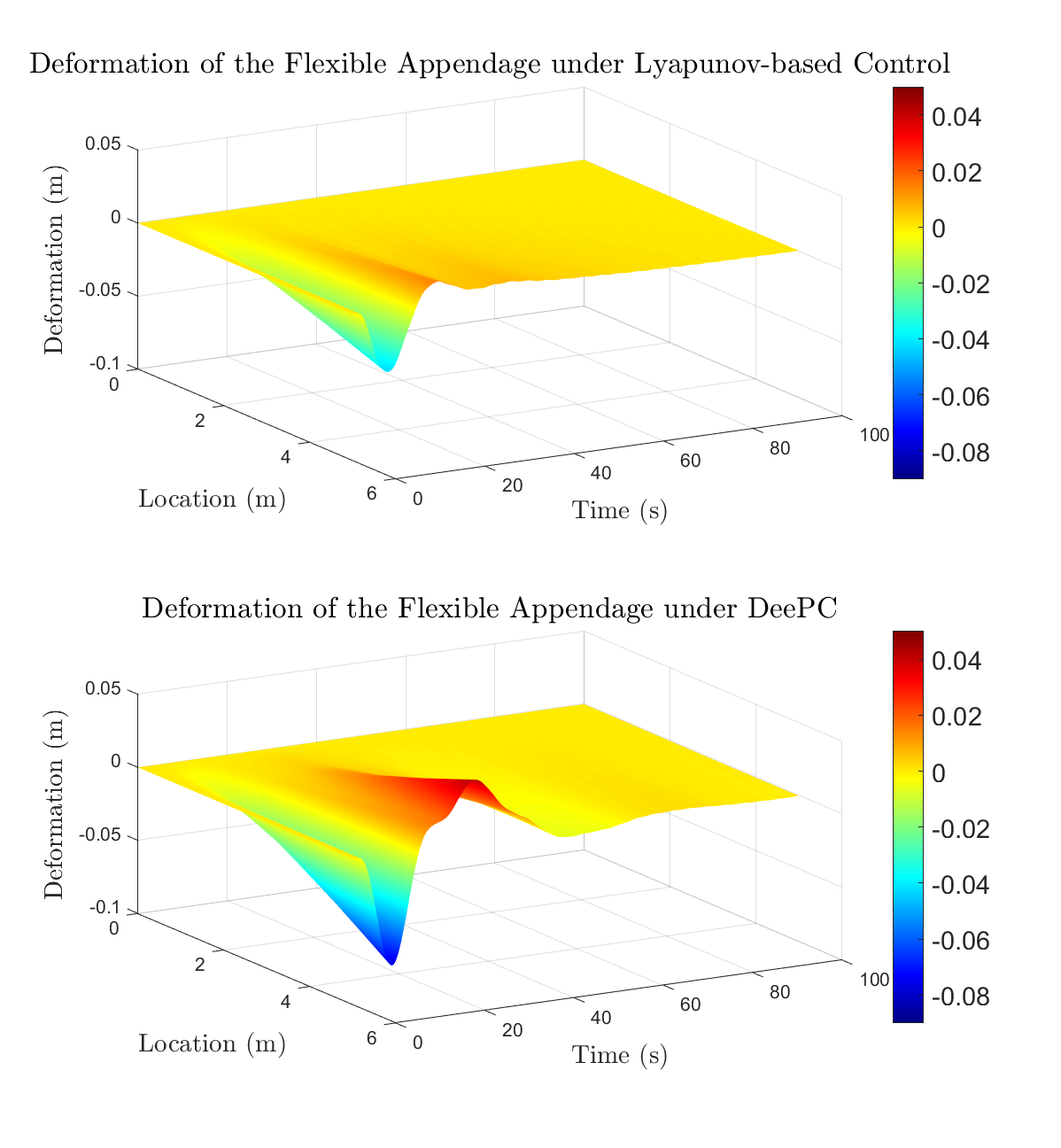}
		\caption{The deformation of the flexible spacecraft with Lyapunov-based control vs DeePC control under model uncertainty. }\label{fig_def2}
	\end{figure}

\begin{table*}[!h]
\centering
\caption{\fontsize{10}{10}\selectfont Lyapunov-based Control vs DeePC under Different Scenarios}
\label{table_example}
\resizebox{\textwidth}{!}{%
\begin{threeparttable}
\setlength{\tabcolsep}{3mm}{
\begin{tabular}{c | c  c  c | c  c}
\hline
\hline
& \multicolumn{3}{c|}{\textbf{Cost Function}} & \multicolumn{2}{c}{\textbf{Settling Time (s)}} \\
& \textbf{Nominal Case} & \textbf{Uncertainty Case} & \textbf{Process Noise Case} & \textbf{Nominal Case} & \textbf{Uncertainty Case} \\
\hline
Lyapunov Method & 2741 & 3870 & 2801 & 58.6 & 90.6 \\
DeePC Method & 2287 & 2623 & 2290 & 62.8 & 64.3 \\
\hline
\hline
\end{tabular}%
}
\end{threeparttable}
}
\end{table*}

A qualitative comparison is presented in Table~\ref{table_example}, where both control methodologies are evaluated based on the cost function across three different scenarios for tracking a reference angle of \( 0.1 \, \mathrm{rad} \) from 0. Additionally, the settling time is compared for the nominal case and the uncertainty case.
As shown in the graphs and the table, DeePC achieves performance that is comparable to or even better than the Lyapunov-based approach, particularly in uncertain cases. It controls the flexible appendage without requiring the derivation or prior knowledge of the system's mathematical model. Its advantage lies in achieving a lower cost function and faster settling time. However, it has larger initial deflection due to its higher control torque, which enables the system to be driven away from undesirable states more rapidly. From an implementation and time-efficiency perspective, developing a parametric model is highly time-consuming, as it requires deriving Hamilton's principles and identifying parameters through experimental data. In contrast, this study presents a systematic approach for applying regularized DeePC, effectively addressing the challenges posed by this complex system.

	\section{Conclusion} \label{sec_conclusion}
 This paper introduced a novel application of the DeePC (data-enabled predictive control) approach for the control of a flexible spacecraft. The study began by providing an overview of the DeePC framework, including its theoretical foundations and key principles, followed by the formulation of the DeePC method specifically tailored to address the challenges associated with flexible spacecraft. Furthermore, the DeePC approach was systematically compared to a well-established Lyapunov-based control method in a finite element (FE) simulation environment. The simulation results provided strong evidence of the validity and effectiveness of the DeePC approach, demonstrating its capability to achieve accurate angle tracking and suppress vibrations without requiring prior knowledge of the system's parameters or the derivation of a mathematical model. This work highlighted DeePC as a powerful and efficient tool for addressing the control challenges of complex systems such as flexible spacecraft.

\appendices
\section{Finite Element Model for the Flexible Spacecraft} \label{sec_appendix_fe}

The coupled PDE-ODE system \eqref{eq:pde:1}-\eqref{eq:pde:4} is numerically solved using a finite element approach. 
First, the standard Euler-Bernoulli beam element \cite{Friedmann2023} is applied to discretize the structural deformation as
\begin{equation}
y(x,t) = \sum_{i=1}^N \phi_i(x)a_i(t) \equiv \phi^\top a,
\end{equation}
where $\phi_i(x)$ are the Hermite shape functions defined on each of the element, and $a_i(t)$ physically represent the displacement and rotation at the nodes of the elements. Next, \eqref{eq:pde:2} is discretized as
\begin{equation}\label{eq:fe:1}
    M a_{tt} + K a = f,
\end{equation}
where mass matrix $M=\int_0^L \rho \phi\phi^\top dx$, stiffness matrix $K=\int_0^L EI \phi_{xx}\phi_{xx}^\top dx$, and the forcing vector is
\begin{equation}\label{eq:fe:2}
    f = -\int_0^L \rho x \phi \theta_{tt} dx \equiv -m\theta_{tt}.
\end{equation}
The geometrical boundary conditions \eqref{eq:pde:3} are satisfied by choosing appropriate shape functions $\phi$, and the natural boundary conditions \eqref{eq:pde:4} are satisfied automatically via the definition of $f$.  Subsequently, \eqref{eq:pde:1} is discretized.  Note that directly discretizing \eqref{eq:pde:1} and coupling it to \eqref{eq:fe:1} would result in a non-symmetric augmented mass matrix; this may cause numerical instabilities in long-term time-accurate solutions.  To avoid such unfavorable property, \eqref{eq:pde:1} is transformed to the following equivalent form \cite{Ge1998},
\begin{equation}\label{eq:pde:5}
    \tilde{J}\theta_{tt}(t) + \rho\int_0^L x y_{tt}(x,t)dx = \tau(t)
\end{equation}
where $\tilde{J}=J+\frac{1}{3}\rho L^3$ is the total moment of inertia of the hub and beam.  The discretized form of \eqref{eq:pde:5} is
\begin{equation}\label{eq:fe:3}
    \tilde{J}\theta_{tt} + m^\top a_{tt} = \tau.
\end{equation}
Combining \eqref{eq:fe:1}, \eqref{eq:fe:2}, and \eqref{eq:fe:3}, the final dynamical model is
\begin{equation}\label{eq:fe:4}
    \begin{bmatrix}
        M & m \\ m^\top & \tilde{J}
    \end{bmatrix}
    \begin{bmatrix}
        a_{tt} \\ \theta_{tt}
    \end{bmatrix}
    +
    \begin{bmatrix}
        K & 0 \\ 0 & 0
    \end{bmatrix}
    \begin{bmatrix}
        a \\ \theta
    \end{bmatrix}
    =
    \begin{bmatrix}
        0 \\ \tau
    \end{bmatrix}.
\end{equation}
Clearly now the augmented mass matrix is real symmetric.  Lastly, \eqref{eq:fe:4} is numerically solved using the Generalized-$\alpha$ method, that is second-order time-accurate and numerically stable for structural dynamic systems \cite{Friedmann2023}.

\section{Proof of Lyapunov Function} \label{sec_appendix_stability}
In order for \(V(t)>0\), where 
\begin{equation}
\begin{aligned}
V(t) =& E_a(t) + \frac{EIa_1}{2} e^2(t) + \frac{1}{2} \Delta^2(t)\\
&\quad + a_3 V_1(t) + a_4 V_2(t),
\end{aligned}
\end{equation}
we will first examine \(V_1\) and \(V_2\).
The following holds for \( V_1 \):

\begin{equation}
\begin{aligned}
    V_1(t) &= \rho \int_0^L (x - L) z_t(x, t) y_x(x, t) \, dx \\
    &\geq -\rho \int_0^L \frac{L - x}{2} \left( z_t^2(x, t) + y_x^2(x, t) \right) dx \\
    &\geq -\frac{\rho L}{2} \int_0^L \left( z_t^2(x, t) + y_x^2(x, t) \right) dx,
\end{aligned}
\end{equation}

where, using Wirtinger's Inequality~\cite{krstic2008}, for any \( \phi \in C^1([0, L]) \),

\begin{equation}
    \int_0^L \phi^2(x) \, dx \leq L \phi^2(0) + \frac{4L^2}{\pi^2} \int_0^L \phi_x^2(x) \, dx,
\end{equation}

we further obtain

\begin{equation}
\begin{aligned}
    \int_0^L y_x^2(x, t) \, dx &\leq L y_x^2(0, t) + \frac{4L^2}{\pi^2} \int_0^L y_{xx}^2(x, t) \, dx  
    \\ &= \frac{4L^2}{\pi^2} \int_0^L y_{xx}^2(x, t) \, dx,
\end{aligned}
\end{equation}

and thus,
\begin{equation}
\begin{aligned}
    V_1(t) &\geq -\frac{\rho L}{2} \int_0^L z_t^2(x, t) \, dx - \frac{\rho L}{2} \int_0^L y_x^2(x, t) \, dx  \\
    &\geq -\frac{\rho L}{2} \int_0^L z_t^2(x, t) \, dx - \frac{2 \rho L^3}{\pi^2} \int_0^L y_{xx}^2(x, t) \, dx.
\end{aligned}
\end{equation}

Similarly,
\begin{equation}
\begin{aligned}
    &V_2(t) = \rho \int_0^L z_t(x, t) \left( xe(t) + y(x, t) \right) \, dx \\
    &\geq -\frac{\rho}{2} \int_0^L \left( z_t^2(x, t) + \left( xe(t) + y(x, t) \right)^2 \right) \, dx \\
    &\geq -\frac{\rho}{2} \int_0^L z_t^2(x, t) \, dx \\
    &- \frac{\rho}{2} \big( L (0 e(t) + y(0, t))^2 + \frac{4L^2}{\pi^2} \int_0^L \big(e(t)+y_x(x, t) \big)^2\, dx \big) \\
    &= -\frac{\rho}{2} \int_0^L z_t^2(x, t) \, dx  \\
    &\quad - \frac{2 \rho L^2}{\pi^2} \int_0^L \left( e(t) + y_x(x, t) \right)^2 \, dx \\
    &\geq -\frac{\rho}{2} \int_0^L z_t^2(x, t) \, dx \\
    &\quad - \frac{2 \rho L^2}{\pi^2} \left( L (e(t) + y_x(0, t))^2 + \frac{4L^2}{\pi^2} \int_0^L y_{xx}^2(x, t) \, dx \right) \\
    &= -\frac{\rho}{2} \int_0^L z_t^2(x, t) \, dx  \\
    &\quad - \frac{2 \rho L^3}{\pi^2} e^2(t) - \frac{8 \rho L^4}{\pi^4} \int_0^L y_{xx}^2(x, t) \, dx.
\end{aligned}
\end{equation}

Therefore,
\begin{equation}
\begin{aligned}
    &V(t) = E_a(t) + \frac{E I a_1}{2} e^2(t) + \frac{1}{2} \Delta^2 + a_3 V_1(t) + a_4 V_2(t) \\
    &\geq \frac{\rho}{2} \int_0^L z_t^2(x, t) \, dx + \frac{E I}{2} \int_0^L y_{xx}^2(x, t) \, dx + \frac{E I a_1}{2} e^2(t)  \\ 
    &\quad + \frac{1}{2} \Delta^2 - \frac{\rho L a_3}{2} \int_0^L z_t^2(x, t) \, dx  \\
    &\quad - \frac{2 \rho L^3 a_3}{\pi^2} \int_0^L y_{xx}^2(x, t) \, dx - \frac{\rho a_4}{2} \int_0^L z_t^2(x, t) \, dx  \\
    &\quad - \frac{2 \rho L^3 a_4}{\pi^2} e^2(t) - \frac{8 \rho L^4 a_4}{\pi^4} \int_0^L y_{xx}^2(x, t) \, dx \\
    &= \frac{\rho}{2} (1 - L a_3 - a_4) \int_0^L z_t^2(x, t) \, dx  \\
    &\quad + \left( \frac{E I}{2} - \frac{2 \rho L^3 a_3}{\pi^2} - \frac{8 \rho L^4 a_4}{\pi^4} \right) \int_0^L y_{xx}^2(x, t) \, dx  \\
    &\quad + \left( \frac{E I a_1}{2} - \frac{2 \rho L^3 a_4}{\pi^2} \right) e^2(t) + \frac{1}{2} \Delta^2.
\end{aligned}
\end{equation}

We consider $(a_1, a_2, a_3, a_4)$ satisfying the following constraints:

\begin{align}
    &a_1, a_2, a_3, a_4 > 0, \\
    &1 - L a_3 - a_4 > 0, \\
    &\frac{E I}{2} - \frac{2 \rho L^3 a_3}{\pi^2} - \frac{8 \rho L^4 a_4}{\pi^4} > 0, \\
    &\frac{E I a_1}{2} - \frac{2 \rho L^3 a_4}{\pi^2} > 0,
\end{align}
which, for any given \( a_1, a_2 > 0 \), can be satisfied for sufficiently small \( a_3, a_4 > 0 \).

Under the above constraints, \( V \geq 0 \) and \( V = 0 \) if and only if \( e = 0 \) \(\iff\) \( \theta = \theta_d, z_t = y_{xx} = 0 \) on \( [0, L] \) almost everywhere, and \( J \theta_t - a_2 y_{xx}(0) = 0 \).
We only consider classical/smooth solutions, thus \( y_{xx} = 0 \) on \( [0, L] \) everywhere \(\implies y_{xx}(0) = 0 \implies \theta_t = 0 \).


We now consider the time derivative of \( V(t) \).

Firstly, we investigate the components of \( \frac{d}{dt} E_a(t) \). As preliminaries, the following holds:
\begin{equation}
\begin{aligned}
    &\frac{d}{dt} \frac{\rho}{2} \int_0^L z_t^2(x, t) \, dx = \rho \int_0^L z_t(x, t) z_{tt}(x, t) \, dx \\
    &= - E I \int_0^L \left( x \theta_t(t) + y_t(x,t)  \right) y_{xxxx}(x, t) \, dx \\
    &= - E I \theta_t(t) \int_0^L x \, dy_{xxx}(x, t)  
    - E I \int_0^L y_{t}(x, t) \, dy_{xxx}(x, t) \\
    &= - E I \theta_t(t) \big( L y_{xxx}(L, t) - 0 y_{xxx}(0, t)  
    - \int_0^L y_{xxx}(x, t) \, dx \big)\\
    &\quad - E I \big( y_t(L, t) y_{xxx}(L, t) - y_t(0, t) y_{xxx}(0, t) \\
    &\quad \quad - \int_0^L y_{xxx}(x, t) \, dy_t(x, t) \big) \notag \\
\end{aligned}
\end{equation}
\begin{equation}
\begin{aligned}
    &= E I \theta_t(t) \int_0^L y_{xxx}(x, t) \, dx + E I \int_0^L y_{xxx}(x, t) dy_t(x, t) \\
    &= E I \theta_t(t) (y_{xx}(L, t) - y_{xx}(0, t))  \\
    &\quad + E I \int_0^L y_{xt}(x, t) y_{xxx}(x, t) \, dx \\
    &= - E I \theta_t(t) y_{xx}(0, t) + E I \int_0^L y_{xt}(x, t) dy_{xx}(x, t) \\
    &= - E I \theta_t(t) y_{xx}(0, t)  \\
    &\quad + E I \big( y_{xt}(L, t) y_{xx}(L, t) - y_{xt}(0, t) y_{xx}(0, t)   \\
    &\quad - \int_0^L y_{xx}(x, t) dy_{xt}(x, t) \big) \\
    &= - E I \theta_t(t) y_{xx}(0, t) - E I \int_0^L y_{xx}(x, t) dy_{xt}(x, t),
\end{aligned}
\end{equation}

for the set of solutions satisfying

\begin{equation}
    \frac{d}{dt} \int_0^L z_t^2(x, t) \, dx = \int_0^L \frac{d}{dt} z_t^2(x, t) \, dx.
\end{equation}

Thus,
\begin{equation}
\begin{aligned}
    \frac{d}{dt} E_a(t) &= - E I \theta_t(t) y_{xx}(0, t) - E I \int_0^L y_{xx}(x, t) dy_{xt}(x, t)  \\
    &\quad + \frac{d}{dt} \frac{E I}{2} \int_0^L y_{xx}^2(x, t) \, dx \\
    &= - E I \theta_t(t) y_{xx}(0, t) - E I \int_0^L y_{xx}(x, t) dy_{xt}(x, t)  \\
    &\quad + E I \int_0^L y_{xx}(x, t) y_{xxt}(x, t) \, dx\\
    &= - E I \theta_t(t) y_{xx}(0, t) ,
\end{aligned}
\end{equation}

for the set of solutions satisfying
\begin{align}
    \frac{d}{dt} \int_0^L y_{xx}^2(x, t) \, dx = \int_0^L \frac{d}{dt} y_{xx}^2(x, t) \, dx.
\end{align}

Then, we work through \(\frac{d}{dt} \Delta(t) \):
\begin{equation}
\begin{aligned}
    \frac{d}{dt} \Delta(t) &= J \theta_{tt}(t) - a_2 y_{xxt}(0, t) + a_1 a_2 e_t(t) \\
    &= \tau(t) + E I y_{xx}(0, t) - a_2 y_{xxt}(0, t) \\
    &\quad + a_1 a_2 e_t(t).
\end{aligned}
\end{equation}

We now consider the time derivative of \( V_1(t) \).

\begin{equation}
\begin{aligned}
    \frac{d}{dt} V_1(t) &= \rho \int_0^L (x - L) z_{tt}(x, t) y_x(x, t) \, dx \\
    &\quad + \rho \int_0^L (x - L) z_t(x, t) y_{xt}(x, t) \, dx \\
    &= - E I \int_0^L (x - L) y_{xxxx}(x, t) y_x(x, t) \, dx  \\
    &\quad + \rho \int_0^L (x - L) z_t(x, t) y_{xt}(x, t) \, dx.
\end{aligned}
\end{equation}

To simplify the first component of \(\frac{d}{dt} V_1(t)\) from the equation above, the following preliminaries hold:
\begin{equation}
\begin{aligned}
    &\int_0^L (x - L) y_{xxxx}(x, t) y_x(x, t) \, dx \\
    &= \int_0^L (x - L) y_x(x, t) \, dy_{xxx}(x, t) \\
    &= L y_x(0, t) y_{xxx}(0, t) - \int_0^L y_{xxx}(x, t) d((x - L) y_x(x, t)) \\
    &= - \int_0^L (x - L) y_{xxx}(x, t) dy_x(x, t)  \\
    &\quad - \int_0^L y_{xxx}(x, t) y_{x}(x, t) \, dx \\
    &= - \int_0^L (x - L) y_{xx}(x, t) dy_{xx}(x, t) \\
    &\quad - \int_0^L y_x(x, t) dy_{xx}(x, t)  \\
    &= - \frac{1}{2} \int_0^L (x - L) d(y_{xx}(x, t))^2  \\
    &\quad - \bigg( y_x(L, t) y_{xx}(L, t) - y_x(0, t) y_{xx}(0, t)   \\
    &\quad - \int_0^L y_{xx}(x, t) dy_x(x, t) \bigg) \\
    &= - \frac{1}{2} ( L (y_{xx}(0, t))^2 - \int_0^L (y_{xx}(x, t))^2 \, dx ) \\
    &\quad + \int_0^L (y_{xx}(x, t))^2 \, dx \\
    &= - \frac{L}{2}  y_{xx}^2(0, t) + \frac{3}{2} \int_0^L y_{xx}^2(x, t) \, dx.
\end{aligned}
\end{equation}

And the second component of \(\frac{d}{dt} V_1(t) \) is worked out as follows:
\begin{equation}
\begin{aligned}
    &\int_0^L (x - L) z_t(x, t) y_{xt}(x, t) \, dx \\
    &= \int_0^L (x - L) (x \theta_t(t) + y_t(x, t)) y_{xt}(x, t) \, dx \\
    &= \theta(t) \int_0^L (x^2 - L x) y_{xt}(x, t) \, dx  \\
    &\quad + \int_0^L (x - L) y_t(x, t) y_{xt}(x, t) \, dx \\
    &= \theta(t) \int_0^L (x^2 - L x) \, dy_t(x, t)  \\
    &\quad + \int_0^L (x - L) y_t(x, t) \, dy_t(x, t) \\
    &= - \theta(t) \int_0^L y_t(x, t) d(x^2 - L x)  \\
    &\quad + \frac{1}{2} \int_0^L (x - L) d (y_t^2(x, t)) \notag \\
\end{aligned}
\end{equation}

\begin{equation}
\begin{aligned}    
    &= - \theta_t(t) \int_0^L (2x - L) y_t(x, t) \, dx - \frac{1}{2} \int_0^L y_t^2(x, t) \, dx \\
    &\leq \int_0^L \frac{L}{2} (\frac{1}{\delta} \theta_t^2(t) + \delta y_t^2(x, t)) \, dx - \frac{1}{2} \int_0^L y_t^2(x, t) \, dx \\
    &= \frac{L^2}{2 \delta} \theta_t^2(t) + \frac{L \delta}{2} \int_0^L y_t^2(x, t) \, dx - \frac{1}{2} \int_0^L y_t^2(x, t) \, dx \\
    &= \frac{L^2}{2 \delta} \theta_t^2(t) - \frac{1}{2} (1 - L \delta) \int_0^L y_t^2(x, t) \, dx.
\end{aligned}
\end{equation}
Thus, we have
\begin{equation}
\begin{aligned}
    \frac{d}{dt} V_1(t) &= - E I \int_0^L (x - L) y_{xxxx}(x, t) y_x(x, t) \, dx \\
    &\quad + \rho \int_0^L (x - L) z_t(x, t) y_{xt}(x, t) \, dx \\
    &\leq \frac{L E I}{2} y_{xx}^2(0, t) - \frac{3 E I}{2} \int_0^L y_{xx}^2(x, t) \, dx  \\
    &\quad + \frac{\rho L^2}{2 \delta} \theta_t^2(t) - \frac{\rho}{2} (1 - L \delta) \int_0^L y_t^2(x, t) \, dx.
\end{aligned}
\end{equation}
We now consider the time derivative of \( V_2(t) \):
\begin{equation}
\begin{aligned}
    \frac{d}{dt} V_2(t) &= \rho \int_0^L z_{tt}(x, t) (x e(t) + y(x, t)) \, dx  \\
    &\quad + \rho \int_0^L z_t(x, t) (x e(t) + y_t(x, t)) \, dx \\
    &= \rho \int_0^L z_{tt}(x, t) (x e(t) + y(x, t)) \, dx  \\
    &\quad + \rho \int_0^L z_t^2(x, t) \, dx \\
    &= - E I \int_0^L y_{xxxx}(x, t) (x e(t) + y(x, t)) \, dx  \\
    &\quad + \rho \int_0^L z_t^2(x, t) \, dx,
\end{aligned}
\end{equation}
where
\begin{equation}
\begin{aligned}
    &\int_0^L y_{xxxx}(x, t) (x e(t) + y(x, t)) \, dx \\
    &= e(t) \int_0^L x y_{xxxx}(x, t) \, dx + \! \int_0^L y(x, t) y_{xxxx}(x, t) \, dx \\
    &= e(t) \int_0^L x \, dy_{xxx}(x, t) + \int_0^L y(x, t) \, dy_{xxx}(x, t) \\
    &= e(t) \left( L y_{xxx}(L, t) - \int_0^L y_{xxx}(x, t) \, dx \right) \\
    &\quad + \bigg( y(L, t) y_{xxx}(L, t) - y(0, t) y_{xxx}(0, t)  \\
    &\quad - \int_0^L y_{xxx}(x, t) dy(x, t) \bigg)\\
    &= - e(t) \int_0^L dy_x(x, t) - \int_0^L y_x(x, t) y_{xxx}(x, t) \, dx \\
    &= - e(t) (y_{xx}(L, t) - y_{xx}(0, t)) - \int_0^L y_x(x, t) dy_{xx}(x, t) \\
    &= e(t) y_{xx}(0, t) - \bigg( y_x(L, t) y_{xx}(L, t) - y_x(0, t) y_{xx}(0, t)   \\
    &\quad - \int_0^L y_{xx}(x, t) dy_x(x, t) \bigg)\\
    &= e(t) y_{xx}(0, t) + \int_0^L y_{xx}^2(x, t) \, dx,
\end{aligned}
\end{equation}

and
\begin{equation}
\begin{aligned}
    \int_0^L z_t^2(x, t) \, dx &= \int_0^L (x \theta_t(t) + y_t(x, t))^2 \, dx \\
    &\leq \int_0^L 2x^2 \theta_t^2(t) + 2y_t^2(x, t) \, dx \\
    &= \frac{2L^3}{3} \theta_t^2(t) + 2 \int_0^L y_t^2(x, t) \, dx.
\end{aligned}
\end{equation}

Thus,
\begin{equation}
\begin{aligned}
    \frac{d}{dt} V_2(t) &= -E I \int_0^L y_{xxxx}(x, t) (x e(t) + y(x, t)) \, dx \\
    &\quad + \rho \int_0^L z_t^2(x, t) \, dx \\
    &= -E I e(t) y_{xx}(0, t) - E I \int_0^L y_{xx}^2(x, t) \, dx \\
    &\quad + \rho \int_0^L z_t^2(x, t) \, dx \\
    &\leq -E I e(t) y_{xx}(0, t) - E I \int_0^L y_{xx}^2(x, t) \, dx \\
    &\quad + \frac{2 \rho L^3}{3} \theta_t^2(t) + 2 \rho \int_0^L y_t^2(x, t) \, dx.
\end{aligned}
\end{equation}

Therefore, we have
\begin{equation}
\begin{aligned}
    \frac{d}{dt} V(t) &= \frac{d}{dt} E_a(t) + E I a_1 e(t) e_t(t) + \Delta(t) \frac{d}{dt} \Delta(t) \\
    &\quad + a_3 \frac{d}{dt} V_1(t) + a_4 \frac{d}{dt} V_2(t) \\
    &\leq - E I \theta_t(t) y_{xx}(0, t) + E I a_1 e(t) \theta_t(t) \\
    &\quad + \Big( J \theta_t(t) + a_2 \big( - y_{xx}(0, t) + a_1 e(t) \big) \Big) \\
    &\quad \times \! \Big( \tau(t) + E I y_{xx}(0, t) - a_2 y_{xxt}(0, t) + a_1 a_2 e_t(t) \Big) \\
    &\quad + a_3 \Big( \frac{L E I}{2} y_{xx}^2(0, t) - \frac{3 E I}{2} \int_0^L y_{xx}^2(x, t) \, dx  \\
    &\quad + \frac{\rho L^2}{2 \delta} \theta_t^2(t) - \frac{\rho}{2} (1 - L \delta) \int_0^L y_t^2(x, t) \, dx \Big) \\
    &\quad + a_4 \Big( - E I e(t) y_{xx}(0, t) - E I \int_0^L y_{xx}^2(x, t) \, dx \\
    &\quad + \frac{2 \rho L^3}{3} \theta_t^2(t) + 2 \rho \int_0^L y_t^2(x, t) \, dx \Big) \\
    &= E I \Big( - y_{xx}(0, t) + a_1 e(t) \Big) \theta_t(t) \\
    &\quad + \! \Big( \tau(t) + E I y_{xx}(0, t) - a_2 y_{xxt}(0, t) + a_1 a_2 \theta_t(t) \Big) \\
    &\quad \times \Big( J \theta_t(t) + a_2 \big( - y_{xx}(0, t) + a_1 e(t) \big) \Big) \dots 
\end{aligned}
\end{equation}

We now consider the following control:
\begin{equation}
\begin{aligned}
    \tau(t) &= \left( - E I y_{xx}(0, t) + a_2 y_{xxt}(0, t) - a_1 a_2 \theta_t(t) \right) -  \frac{E I}{a_2} \theta_t(t)\\
    &\quad  - k_1 \Delta(t) + k_2 \left( J \theta_t(t) + a_2 y_{xx}(0, t) - a_1 a_2 e(t) \right).
\end{aligned}
\end{equation}

Then, it holds that
\begin{equation}
\begin{aligned}
    \frac{d}{dt} V(t) &\leq E I \big( -y_{xx}(0, t) + a_1 e(t) \big) \theta_t(t) + \\
    &\quad \big( \tau(t) + E I y_{xx}(0, t) - a_2 y_{xxt}(0, t) + a_1 a_2 \theta_t(t) \big) \\
    &\quad \times \big( J \theta_t(t) + a_2 \big( -y_{xx}(0, t) + a_1 e(t) \big) \big)  \\
    &\quad + a_3 \bigg( \frac{L E I}{2} y_{xx}^2(0, t) - \frac{3 E I}{2} \int_0^L y_{xx}^2(x, t) dx \\
    &\quad+ \frac{\rho L^2}{2 \delta} \theta_t^2(t) - \rho \frac{(1 - L \delta)}{2} \int_0^L y_t^2(x, t) dx \bigg) \\
    &\quad + a_4 \bigg( -E I e(t) y_{xx}(0, t) - E I \int_0^L y_{xx}^2(x, t) dx \\
    &\quad + \frac{2 \rho L^3}{3} \theta_t^2(t) + 2 \rho \int_0^L y_t^2(x, t) dx \bigg) \\
\end{aligned}
\end{equation}

\begin{equation}
\begin{aligned}
    &= E I \big( -y_{xx}(0, t) + a_1 e(t) \big) \theta_t(t) + \\
    &\quad \Big( -\frac{E I}{a_2} \theta_t(t) - k_1 \Delta(t) + k_2 \big( J \theta_t(t) + a_2 y_{xx}(0, t)\\
    &\quad- a_1 a_2 e(t) \big) \Big) \\
    &\quad \times \Big( J \theta_t(t) + a_2 \big( -y_{xx}(0, t) + a_1 e(t) \big) \Big) + \\
    &\quad a_3 \Big( \frac{L E I}{2} y_{xx}^2(0, t) - \frac{3 E I}{2} \int_0^L y_{xx}^2(x, t) \, dx \\
    &\quad + \frac{\rho L^2}{2 \delta} \theta_t^2(t) - \frac{\rho}{2} (1 - L \delta) \int_0^L y_t^2(x, t) \, dx \Big) + \\
    &\quad a_4 \Big( -E I e(t) y_{xx}(0, t) - E I \int_0^L y_{xx}^2(x, t) \, dx \\
    &\quad + \frac{2 \rho L^3}{3} \theta_t^2(t) + 2 \rho \int_0^L y_t^2(x, t) \, dx \Big)
\end{aligned}
\end{equation}


\begin{equation}
\begin{aligned}
    &=- k_1 \Delta^2(t) - \frac{J E I}{a_2} \theta_t^2(t) + k_2 J^2 \theta_t^2(t) - k_2 a_2^2 \big( y_{xx}(0, t) \\
    &\quad - a_1 e(t) \big)^2+ \frac{L E I a_3}{2} y_{xx}^2(0, t) \\
    &\quad + \big( \frac{\rho L^2 a_3}{2 \delta} + \frac{2 \rho L^3 a_4}{3} \big) \theta_t^2(t) - E I a_4 e(t) y_{xx}(0, t) \\
    &\quad - E I \left( \frac{3 a_3}{2} + a_4 \right) \int_0^L y_{xx}^2(x, t) \, dx \\
    &\quad- \rho \left( \frac{a_3}{2} (1 - L \delta) - 2 a_4 \right) \int_0^L y_t^2(x, t) \, dx \\
    &= -k_1 \Delta^2(t) - \big( \frac{J E I}{a_2} - k_2 J^2 \\
    &\quad- \frac{\rho L^2 a_3}{2 \delta} - \frac{2 \rho L^3 a_4}{3} \big) \theta_t^2(t) - \big( k_2 a_2^2 - \frac{L E I a_3}{2} \big) y_{xx}^2(0, t) \\
    &\quad - k_2 a_1^2 a_2^2 e^2(t) + \left( 2 k_2 a_1 a_2^2 - E I a_4 \right) e(t) y_{xx}(0, t) \\
    &\quad - E I \big( \frac{3 a_3}{2} + a_4 \big) \int_0^L y_{xx}^2(x, t) \, dx \\
    &\quad- \rho \big( \frac{a_3}{2} (1 - L \delta) - 2 a_4 \big) \int_0^L y_t^2(x, t) \, dx. 
\end{aligned}
\end{equation}

We look at the terms:

\begin{equation}
\begin{aligned}
    &- \left( k_2 a_2^2 - \frac{L E I a_3}{2} \right) y_{xx}^2(0, t) - k_2 a_1^2 a_2^2 e^2(t) \\
    & + \big( 2 k_2 a_1 a_2^2 - E I a_4 \big) e(t) y_{xx}(0, t) \\
    &\leq - \left( k_2 a_2^2 - \frac{L E I a_3}{2} \right) y_{xx}^2(0, t) - k_2 a_1^2 a_2^2 e^2(t) \\
    &\quad + \left( k_2 a_1 a_2^2 - \frac{E I a_4}{2} \right) \left( \frac{1}{\eta} e^2(t) + \eta y_{xx}^2(0, t) \right) \\
    &= - \left( k_2 a_2^2 - \eta k_2 a_1 a_2^2 + \frac{E I \eta a_4}{2} - \frac{L E I a_3}{2} \right) y_{xx}^2(0, t) \\
    &\quad - \left( k_2 a_1^2 a_2^2 - \frac{k_2 a_1 a_2^2}{\eta} - \frac{E I a_4}{2 \eta} \right) e^2(t).
\end{aligned}
\end{equation}

We consider $(a_1, a_2, a_3, a_4, k_1, k_2, \delta, \eta)$ satisfying the following constraints:
    
\begin{equation}
\begin{aligned}
    &a_1, a_2, a_3, a_4, k_1, k_2, \delta, \eta > 0, \\
    &\frac{J E I}{a_2} - k_2 J^2 - \frac{\rho L^2 a_3}{2 \delta} - \frac{2 \rho L^3 a_4}{3} > 0, \\
    &(1 - L \delta) a_3 - 4 a_4 > 0,  \\
    &k_2 a_2^2 - \eta k_2 a_1 a_2^2 + \frac{E I \eta a_4}{2} - \frac{L E I a_3}{2} \geq 0,  \\
    &k_2 a_1^2 a_2^2 - \frac{k_2 a_1 a_2^2}{\eta} + \frac{E I a_4}{2 \eta} > 0. 
\end{aligned}
\end{equation}

Under the above constraints, \( \frac{d}{dt} V(t) \leq 0 \) and \( \frac{d}{dt} V(t) = 0 \) if and only if \( e = 0 \iff \theta = \theta_d, \, \theta_t = 0, \, y_t = y_{xx} = 0 \) on \([0, L]\) almost everywhere, and \( \Delta = 0 \implies y_{xx}(0) = 0 \). We only consider classical/smooth solutions, thus, \( y_t = y_{xx} = 0 \) on \([0, L]\) everywhere.

It remains to show that \( y_{xx} = 0 \) on \([0, L]\) everywhere implies \( y = 0 \) on \([0, L]\) everywhere. We use the Wirtinger’s Inequality twice to prove this as follows:
\begin{equation}
\begin{aligned}
    \int_0^L y^2(x, t) dx &\leq L y^2(0, t) + \frac{4L^2}{\pi^2} \int_0^L y_x^2(x, t) dx \\
    &\leq \frac{4L^2}{\pi^2} \left(L y_x^2(0, t) + \frac{4L^2}{\pi^2} \int_0^L y_{xx}^2(x, t) dx \right) \\
    &= \frac{16L^4}{\pi^4} \int_0^L y_{xx}^2(x, t) dx = 0.
\end{aligned}
\end{equation}

Therefore, the Lyapunov functional candidate \( V \) and the selected control \( \tau \) verify the asymptotic convergence of \( \theta(t) \) to \( \theta_d \) and \( \|y(\cdot, t)\|_{L^2} \to 0 \) (indeed, for classical/smooth solutions, we further have \( \|y(\cdot, t)\|_{L^\infty} \to 0 \)).

Finally, we argue that there exist \( (a_1, a_2, a_3, a_4, k_1, k_2, \delta, \eta) \) that satisfy all of the above constraints:

\begin{equation}
\begin{aligned}
    &a_1, a_2, a_3, a_4, k_1, k_2, \delta,\eta > 0,  \\
    &1 - L a_3 - a_4 > 0,  \\
    &\frac{E I}{2} - \frac{2 \rho L^3 a_3}{\pi^2} - \frac{8 \rho L^4 a_4}{\pi^4} > 0,  \\
    &\frac{E I a_1}{2} - \frac{2 \rho L^3 a_4}{\pi^2} > 0,  \\
    &\frac{J E I}{a_2} - k_2 J^2 - \frac{\rho L^2 a_3}{2 \delta} - \frac{2 \rho L^3 a_4}{3} > 0,  \\
    &(1 - L \delta) a_3 - 4 a_4 > 0,  \\
    &k_2 a_2^2 - \eta k_2 a_1 a_2^2 + \frac{E I \eta a_4}{2} - \frac{L E I a_3}{2} \geq 0, \\
    &k_2 a_1^2 a_2^2 - \frac{k_2 a_1 a_2^2}{\eta} + \frac{E I a_4}{2 \eta} > 0. 
\end{aligned}
\end{equation}

To show this, following \cite{RAD2018Acta}, we consider the subset of solutions satisfying
\begin{equation}
    2 k_2 a_1 a_2^2 - E I a_4 = 0 \iff a_4 = \frac{2 k_2 a_1 a_2^2}{E I}, 
\end{equation}
which simplifies the constraints to:
\begin{equation}
\begin{aligned}
    &a_1, a_2, a_3, a_4, k_1, k_2, \delta > 0,  \\
    &1 - L a_3 - a_4 > 0,  \\
    &\frac{E I}{2} - \frac{2 \rho L^3 a_3}{\pi^2} - \frac{8 \rho L^4 a_4}{\pi^4} > 0,  \\
    &\frac{E I a_1}{2} - \frac{2 \rho L^3 a_4}{\pi^2} > 0,  \\
    &\frac{J E I}{a_2} - k_2 J^2 - \frac{\rho L^2 a_3}{2 \delta} - \frac{2 \rho L^3 a_4}{3} > 0,  \\
    &(1 - L \delta) a_3 - 4 a_4 > 0,  \\
    &k_2 a_2^2 - \frac{L E I a_3}{2} \geq 0.
\end{aligned}
\end{equation}

We further let
\begin{equation}
    k_2 a_2^2 - \frac{L E I a_3}{2} = 0 \iff a_3 = \frac{2 k_2}{L E I} a_2^2. 
\end{equation}

Then,
\begin{equation}
\begin{aligned}
    a_4 &< 1 - L a_3 = 1 - \frac{2k_2}{E I} a_2^2, \\
    a_4 & < \frac{\pi^4 E I}{16 \rho L^4} - \frac{\pi^2}{4 L}
    a_3 = \frac{\pi^4 E I}{16 \rho L^4} - \frac{\pi^2 k_2}{2 L^2 E I} a_2^2, \\
    a_4 &< \frac{\pi^2 E I}{4 \rho L^3} a_1, \\
    a_4 &< \frac{3 J E I}{2 \rho L^3 a_2} - \frac{3 J^2 k_2}{2 \rho L^3} - \frac{3}{4 L \delta} a_3 = \frac{3 J  E I}{2 \rho L^3 a_2}-\frac{3 J^2 k_2}{2 \rho L^3} \\
    &\quad - \frac{3 k_2}{2 L^2 E I \delta} a_2^2, \\
    a_4 &< \frac{1 - L \delta}{4} a_3 = \frac{(1 - L \delta) k_2}{2 L E I} a_2^2.
\end{aligned}
\end{equation}

For any \( \delta \in (0, 1/L) \), let
\begin{equation}
    a_1 = \frac{1 - L \delta}{4 L} \varepsilon_1, \quad \varepsilon_1 \in (0, 1). 
\end{equation}

Then, for any \( a_2 > 0 \), let
\begin{equation}
\begin{aligned}
    k_2 = &\varepsilon_2 \min \Bigg( \frac{E I}{2 a_2^2 (a_1 + 1)}, \frac{\pi^4 (E I)^2}{8 \rho L^2 a_2^2 (4 L^2 a_1 + \pi^2)}, \\
    &\frac{\pi^2 (E I)^2}{8 \rho L^3 a_2^2}, \frac{3 J (E I)^2 \delta}{a_2 (4 \rho L^3 \delta a_1 a_2^2 + 3 E I J^2 \delta + 3 \rho L a_2^2)} \Bigg), \\
    & \varepsilon_2 \in (0, 1). 
\end{aligned}
\end{equation}

The qualifying parameters satisfying the Lyapunov function can then be selected following the above simplified requirements.

	
	
	%

	\balance	
	\bibliographystyle{IEEEtran}
	\bibliography{IEEEabrv,reference}
\end{document}